\documentclass[12pt]{article}

\usepackage[utf8]{inputenc}
\usepackage[english]{babel}
\usepackage{amsmath}
\usepackage{amssymb}
\usepackage{bbm}
\usepackage{hyperref}
\usepackage{physics}
\usepackage{graphicx}

\usepackage{cite}

\usepackage{color}

\newcommand{\be}{\begin{equation}}
\newcommand{\ee}{\end{equation}}
\newcommand{\ba}{\begin{eqnarray}}
\newcommand{\ea}{\end{eqnarray}}
\newcommand{\bs}{\begin{subequations}}
\newcommand{\es}{\end{subequations}}
\newcommand{\no}{\nonumber\\}

\graphicspath{ {figures/} }

\begin{document}

\title{\LARGE On the addition of an $SU(2)$ quadruplet of scalars \\
  to the Standard Model}

\author{
  Darius~Jur\v{c}iukonis$^{(1)}$\thanks{E-mail:
    \tt darius.jurciukonis@tfai.vu.lt}
  \
  and Lu\'\i s~Lavoura$^{(2)}$\thanks{E-mail:
    \tt balio@cftp.tecnico.ulisboa.pt}
  \\*[3mm]
  $^{(1)}\!$
  \small Vilnius University, Institute of Theoretical Physics and Astronomy, \\
  \small Saul\.etekio~av.~3, Vilnius 10257, Lithuania
  \\*[2mm]
  $^{(2)}\!$
  \small Universidade de Lisboa, Instituto Superior T\'ecnico, CFTP, \\
  \small Av.~Rovisco~Pais~1, 1049-001~Lisboa, Portugal
}

\maketitle

\begin{abstract}
  We consider the extension of the Standard electroweak Model
  through an $SU(2)$ quadruplet of scalars
  with hypercharge either $3/2$ or $1/2$
  (with an additional reflection symmetry in the latter case).
  We establish,
  through \emph{exact analytical equations},
  the boundaries of the phase spaces
  of the gauge-invariant terms that appear
  in the (renormalizable) scalar potentials.
  We devise procedures for the determination
  of necessary and sufficient bounded-from-below conditions
  on those potentials;
  we emphasize that one mostly needs to scan the scalar potential
  over a few \emph{lines},
  instead of \emph{surfaces},
  in order to establish the boundedness-from-below;
  this fact allows one \emph{to reduce by three orders of magnitude
  the computational time} devoted to that establishment.
\end{abstract}

\section{Introduction}

The Standard Model (SM) of the electroweak interactions
has gauge symmetry $SU(2) \times U(1)$.
It has only one $SU(2)$ doublet of scalar fields.
It produces,
upon spontaneous gauge-symmetry breaking and the Higgs mechanism,
one physical scalar boson.
That boson has been discovered in 2012~\cite{Higgs1,Higgs2},
providing a confirmation of the SM;
still,
the self-interactions of that boson~\cite{pdg} remain untested.
Up to now,
no other fundamental boson has yet been discovered,
despite some hints~\cite{hints1, hints2}.

There is no reason why there should be only one doublet of scalars
in an $SU(2) \times U(1)$ gauge theory.
Physicists have entertained thoughts
that there may be more $SU(2)$ doublets,
and possibly also singlets---preserving the so-called custodial symmetry
of the SM,
\textit{i.e.}\ the tree-level relation $m_W = m_Z \cos{\theta_w}$,
where $m_W$ and $m_Z$ are the masses of the gauge bosons $W^\pm$ and $Z^0$,
respectively,
and $\theta_w$ is the weak mixing (or Weinberg) angle.
One may consider larger $SU(2)$ multiplets,
like triplets~\cite{triplet1,triplet2,triplet3,triplet4},
quadruplets~\cite{kannike},
and more~\cite{we,meloni1,meloni2},
even though they would violate the custodial symmetry if they acquired
vacuum expectation values (VEVs).
Large $SU(2)$ scalar multiplets
bring about complicated scalar potentials (SPs)
with several $SU(2)$-invariant terms,
and those terms may lead to phase spaces\footnote{Some physicists
call them instead `orbit spaces'.
In this context, this concept was first introduced in Ref.~\cite{kim}.}
with unexpectedly complex shapes~\cite{orbit1,orbit2}.
(The same holds true if the gauge group
is chosen to be larger than $SU(2) \times U(1)$~\cite{other1,other2}.)
The exploration of those shapes is relevant since SPs
must be bounded from below (BFB),
lest the theory has no vacuum state.
In order to establish that an SP is BFB
one must carefully scrutinize all its phase space,
and just that phase space.

In this paper we deal on the addition of one $SU(2)$ quadruplet of scalars
to the SM.
In order to bring about interesting possibilities
for its interaction with the doublet of the SM,
we consider two possibilities
for their relative hypercharges\footnote{In a recent paper
of ours~\cite{milagre} the hypercharge of the quadruplet was left arbitrary,
preventing additional interactions with the doublet.}:
either they are equal
or the hypercharge of the quadruplet is three times the one of
the doublet.
The purpose of this paper is to show that,
even in these very simple extensions of the scalar sector of the SM,
phase spaces with complicated shapes arise;
fortunately,
however,
those shapes may be described by \emph{exact analytical equations}.
The establishment of this fact and of those equations
constitute the groundbreaking achievement of this work.

It should be mentioned that scalar quadruplets
like the ones that we envisage in this paper,
\textit{viz.}\ with hypercharge either $1/2$ or $3/2$,
may be useful---when accompanied by large fermionic $SU(2)$ multiplets---to
construct effective extensions of the usual seesaw mechanisms
for the neutrino masses~\cite{meloni1}.
Extensions of the scalar sector of the SM will also bring about
deviations of the self-interactions of the SM's scalar boson
from their SM values,
and it is relevant to check how large such deviations may be~\cite{kannike}.
Scalars in large $SU(2)$ representations are also useful to render
the coupling of the Higgs boson to a pair of gauge bosons larger
than what it is in the SM, if the new scalars are allowed to have
VEVs~\cite{Yagyu1}.\footnote{See also
Ref.~\cite{Yagyu2}.} Those VEVs do not necessarily alter the
$\rho$ parameter~\cite{Yagyu3}.

The plan of this paper is as follows.
In Section~\ref{sec:Notation} we expound relevant material from Ref.~\cite{we}.
In Section~\ref{sec:3/2} we determine the phase space
for a quadruplet with hypercharge $3/2$.
In Section~\ref{sec:1/2} we determine the phase space
for a quadruplet with hypercharge $1/2$ and a reflection symmetry.
In Section~\ref{sec:conclusions} we write down the
main conclusions of this work
and we offer two suggestions for further research.

This paper has many appendices that, however,
do not need to be read in order to understand the relevant computations.
In Appendix~\ref{App:expressions} we display the expressions
of quantities defined
in Sections~\ref{sec:3/2} and~\ref{sec:1/2}.
In Appendix~\ref{App:concavity} we discuss the method to discover whether
a surface is convex,
concave,
or has
undefined concavity properties.
In Appendix~\ref{App:more} we deal on
a further case,
which however is
of no practical interest because it is
unstable under renormalization.
In Appendix~\ref{App:UNI}
(which is not referred to in the main text of the paper)
we display the unitarity constraints on the various models.

\section{Notation}
\label{sec:Notation}

In this section we collect relevant results from Ref.~\cite{we}.
More detailed demonstrations and justifications may be found in that paper.

\paragraph{Multiplets:} Let
\bs
\label{PhiXi}
\ba
\Phi &=& \left( \begin{array}{c} a \\ b \end{array} \right),
\label{Phi}
\\
\Xi &=& \left( \begin{array}{c} c \\ d \\ e \\ f \end{array} \right)
\label{Xi}
\ea
\es
be a doublet and a quadruplet,
respectively,
of the gauge group $SU(2)$.
In Eqs.~\eqref{PhiXi},
$a, \ldots, f$ are complex scalar fields.
Let $I_3$ be the third component of isospin,
then $c$ has $I_3 = 3/2$,
both $a$ and $d$ have $I_3 = 1/2$,
both $b$ and $e$ have $I_3 = -1/2$,
and $f$ has $I_3 = -3/2$.
Let $I_-$ be the lowering operator of isospin,
then $I_- a = b \left/ \sqrt{2} \right.$,
$I_- c = \sqrt{3/2}\, d$,
$I_- d = \sqrt{2}\, e$,
$I_- e = \sqrt{3/2}\, f$,
and $I_- b = I_- f = 0$.
We define $A = \left| a \right|^2$,
$B = \left| b \right|^2$,
$C = \left| c \right|^2$,
$D = \left| d \right|^2$,
$E = \left| e \right|^2$,
and $F = \left| f \right|^2$.
Note that
  \be
  \widetilde \Xi = \left( \begin{array}{c}
    f^\ast \\ - e^\ast \\ d^\ast \\ - c^\ast \end{array} \right)
  \ee
  is also a quadruplet of $SU(2)$,
  and that
  \bs
  \label{tir}
  \ba
  T_1 = \left( \Xi \otimes \Xi \right)_\mathbf{3} &=&
  \sqrt{\frac{1}{5}} \left( \begin{array}{c}
    \sqrt{2} \left( \sqrt{3}\, c e - d^2 \right) \\
    3 c f - d e \\
    \sqrt{2} \left( \sqrt{3}\, d f - e^2 \right)
  \end{array} \right),
  \label{tir1} \\
  T_2 = \left( \Xi \otimes \widetilde \Xi \right)_\mathbf{3} &=&
  \sqrt{\frac{1}{20}} \left( \begin{array}{c}
    \sqrt{6}\, c d^\ast + \sqrt{8}\, d e^\ast + \sqrt{6}\, e f^\ast \\
    3 F + E - D - 3 C \\
    - \sqrt{6}\, c^\ast d - \sqrt{8}\, d^\ast e - \sqrt{6}\, e^\ast f
  \end{array} \right)
  \label{tir2}
  \ea
  \es
  are triplets of $SU(2)$.

\paragraph{Invariants:}
The following quantities are $SU(2)$-invariant:
\bs
\label{thefs}
\ba
F_1 &=& A + B,
\label{f1}
\\
F_2 &=& C + D + E + F,
\label{f2}
\\
F_4 &=& \frac{A - B}{4}\, \left( 3 C + D - E - 3 F \right)
+ \left[ \frac{a b^\ast}{2}\, \left( \sqrt{3}\, c^\ast d + 2 d^\ast e
  + \sqrt{3}\, e^\ast f \right) + \mathrm{H.c.} \right],
\label{f4}
\\
F_5 &=& \frac{2}{5} \left| \sqrt{3}\, c e - d^2 \right|^2
+ \frac{2}{5} \left| \sqrt{3}\, d f - e^2 \right|^2
+ \frac{1}{5} \left| 3 c f - d e \right|^2.
\label{f5}
\ea
\es
The quantities $F_1$,
$F_2$,
and $F_5$ are non-negative,
while $F_4$ may be either positive or negative.
Note that $F_5$ is the norm-squared of the triplet $T_1$ in Eq.~\eqref{tir1}.
All the quantities in Eqs.~\eqref{thefs}
are invariant under separate rephasings of $\Phi$ and $\Xi$,
\textit{i.e.}\ they are $U(1)$-invariant
irrespective of the hypercharges of the two multiplets.

\paragraph{Dimensionless parameters:} We define
\bs
\label{fjo}
\ba
r &=& \frac{F_1}{F_2},
\label{r}
\\
\delta &=& \frac{4 F_4}{3 F_1 F_2},
\label{delta}
\\
\gamma_5 &=& \frac{F_5}{F_2^2}.
\label{g5}
\ea
\es
Both $\delta$ and $\gamma_5$ are invariant
under the re-scalings $\Phi \to \varrho_1 \Phi$ and $\Xi \to \varrho_2 \Xi$,
where $\varrho_1$ and $\varrho_2$ are arbitrary complex numbers;
$r$ is not invariant under those two separate re-scalings,
but it is invariant under the joint re-scaling
$\Phi \to \varrho \Phi,\ \Xi \to \varrho \Xi$,
where $\varrho$ is an arbitrary complex number.

\paragraph{Alternative notation:} In this paragraph
we make a digression from Ref.~\cite{we}
and focus on the notation used in Ref.~\cite{kannike}.
The fields $a, \ldots, f$ are written in that paper as $H_i$
and $\phi_{ijk}$,
where $i, j, k = 1, 2$ and
\bs
\ba
& H_1 = a, \quad H_2 = b, &
\\
& \phi_{111} = c,
\quad
\phi_{112} = \phi_{121} = \phi_{211} = \displaystyle{\frac{d}{\sqrt{3}}},
\quad
\phi_{221} = \phi_{212} = \phi_{122} = \displaystyle{\frac{e}{\sqrt{3}}},
\quad
\phi_{222} = f. &
\ea
\es
Then the quantities
\bs
\label{p1234}
\ba
p_1 &=& \sum_{i=1}^2 \left| H_i \right|^2, \\
p_2 &=& \sum_{i,j,k=1}^2 \left| \phi_{ijk} \right|^2, \\
p_3 &=& \sum_{k,n=1}^2 \Phi_{kn} \Phi_{nk}, \label{7c} \\
p_4 &=& \sum_{k,n=1}^2 \Phi_{kn} H_n^\ast H_k, \label{7d}
\ea
\es
are introduced in Ref.~\cite{kannike}.
In Eqs.~\eqref{7c} and~\eqref{7d},
\be
\Phi_{kn} = \sum_{i,j=1}^2 \phi_{ijk}^\ast \phi_{ijn}.
\ee
Comparing the quantities~\eqref{p1234} with the quantities~\eqref{thefs},
one finds that
\bs
\ba
p_1 &=& F_1, \\
p_2 &=& F_2, \\
p_3 &=& F_2^2 - \frac{10 F_5}{9}, \\
p_4 &=& \frac{F_1 F_2}{2} + \frac{2 F_4}{3}.
\ea
\es
Therefore,
the parameters $\zeta$ and $\xi$ of Ref.~\cite{kannike}
are related to our parameters $\delta$ and $\gamma_5$ through
\bs
\label{zeta}
\ba
\zeta &=& 1 - \frac{10 \gamma_5}{9},
\\
\xi &=& \frac{1+\delta}{2}.
\ea
\es

\paragraph{Gauge $a=0$:}
There is an $SU(2)$ gauge where the field $a$ vanishes in all of space--time.
In that gauge,\footnote{Another interesting gauge is the one where $a = b$
in all of space--time.
We have re-done all our computations using that gauge and we have found
exactly the same results as in the gauge $a = 0$;
in particular,
we have obtained in both gauges
the same expressions for the quantities in Appendix~\ref{App:expressions};
but,
computation times for gauge $a = b$
are significantly larger than for gauge $a = 0$.}
\bs
\ba
F_1 &=& B,
\\
\delta &=& \frac{3 F + E - D - 3 C}{3 \left( F + E + D + C \right)}.
\label{deltaa=0}
\ea
\es

\paragraph{The non-negative quantity $Q_1$:}
We now define
\be
Q_1 = 9 - 9 \delta^2 - 20 \gamma_5.
\label{14a}
\ee
This quantity is non-negative.
This has been demonstrated in Ref.~\cite{we} in a general gauge;
the demonstration is much simpler in the gauge $a=0$,
wherein
\bs
\label{8}
\ba
Q_1 F_2^2 &=& 9 F_2^2 - \left( 3 \delta F_2 \right)^2 - 20 F_5
\\ &=& 9 \left( F + E + D + C \right)^2 - \left( 3 F + E - D - 3 C \right)^2
\no & &
- 8 \left| \sqrt{3}\, ce - d^2 \right|^2
- 8 \left| \sqrt{3}\, df - e^2 \right|^2
- 4 \left| 3 c f - d e \right|^2
\label{8c}
\\ &=& 2 \left| \sqrt{6}\, c^\ast d + \sqrt{8}\, d^\ast e
+ \sqrt{6}\, e^\ast f \right|^2
\\ &\ge& 0.
\ea
\es
Notice that Eq.~\eqref{8} indicates
that the norm-squared of the triplet $T_2$ in Eq.~\eqref{tir2}
is proportional to a linear combination of the $SU(2)$-invariant quatities
$F_2^2$ and $F_5$,
\textit{viz.}\ $9 F_2^2 - 20 F_5$.\footnote{This property may be extended
to the case where $\Xi$ is a larger multiplet of $SU(2)$~\cite{we}.
Namely,
the product $\Xi \times \widetilde \Xi$ always contains a triplet of $SU(2)$,
and the norm-squared of that triplet always is a linear combination of $F_2^2$
(where $F_2$ is the norm-squared of $\Xi$)
and of the norms-squared of all,
except the one with the highest isospin,
$SU(2)$ multiplets in $\Xi \times \Xi$.
Moreover,
the coefficients of those norms-squared have \emph{negative} coefficients,
while $F_2^2$ has positive coefficient,
in the linear combination.}
The inequality $Q_1 \ge 0$
implies that the ranges of the dimensionless parameters~\eqref{fjo}
are $r \in \left[ 0, + \infty \right[$,
$\delta \in \left[ -1, 1 \right]$,
and $\gamma_5 \in \left[ 0, 9/20 \right]$.

\paragraph{Potential:}
The renormalizable and $SU(2) \times U(1)$-invariant scalar potential $V$
of $\Phi$ and $\Xi$ is $V = V_2 + V_4$,
where $V_2 = \mu_1^2 F_1 + \mu_2^2 F_2$
is bilinear in the scalar fields and
\be
V_4 = \frac{\lambda_1}{2}\, F_1^2  + \frac{\lambda_2}{2}\, F_2^2
+ \lambda_3 F_1 F_2 + \lambda_4 F_4 + \lambda_5 F_5 + V_{4,\mathrm{extra}}
\label{V4}
\ee
is quadrilinear in the scalar fields.
In Eq.~\eqref{V4},
$\lambda_1, \lambda_2, \ldots, \lambda_5$ are real dimensionless coefficients;
furthermore,
$V_{4,\mathrm{extra}}$ is a piece of $V_4$
that depends on the specific hypercharge of $\Xi$;
if that hypercharge is neither $0$,
nor $3/2$,
nor $1/2$
then $V_{4,\mathrm{extra}}$ vanishes.\footnote{We assume
a normalization of the hypercharge where $\Phi$
has hypercharge $1/2$.}$^,$\footnote{The case
$V_{4,\mathrm{extra}}=0$ was studied in Ref.~\cite{milagre}.}

\section{Quadruplet with hypercharge $3/2$}
\label{sec:3/2}

\paragraph{Extra term:}
Multiplying $\Phi$ by itself twice
and taking the fully symmetric part of the product,
one sees that
\be
\left( \Phi \otimes \Phi \otimes \Phi \right)_\mathrm{fully\ symmetric}
\propto \left( \begin{array}{c} a^3 \\ \sqrt{3}\, a^2 b \\
  \sqrt{3}\, a b^2 \\ b^3
  \end{array} \right)
\ee
is a quadruplet of $SU(2)$.
Therefore,
if the hypercharge of $\Xi$ is three times the one of $\Phi$,
\textit{i.e.}\ if $\Xi$ has hypercharge $3/2$,
then
\be
%%%%% BEFORE WE USED A PARAMETER \XI. I HAVE NOW NAMED IT \PSI
%%%%% IN ORDER TO AVOID CONFUSION WITH KANNIKE'S \XI OF EQ. (10B).
V_{4,\mathrm{extra}} = \frac{\psi}{2}\, I_{3/2} + \mathrm{H.c.},
\label{VextrY32}
\ee
where $\psi$ is a complex dimensionless coefficient and
\be
I_{3/2} = a^3 c^\ast + \sqrt{3}\, a^2 b d^\ast
+ \sqrt{3}\, a b^2 e^\ast + b^3 f^\ast
\ee
is $SU(2) \times U(1)$-invariant.
We define the dimensionless non-negative real parameter $\epsilon$ as
\be
\epsilon =
\frac{\left| I_{3/2} \right|}{\sqrt{F_1^3 F_2}}.
\label{epsilon}
\ee
Note that $\epsilon$ is invariant
under the re-scalings $\Phi \to \varrho_1 \Phi,\ \Xi \to \varrho_2 \Xi$.
Then,
\be
V_{4,\mathrm{extra}} = \left| \psi \right| \epsilon \, \sqrt{F_1^3 F_2}
\cos \left( \arg \psi + \arg I_{3/2} \right),
\ee
so that
\be
V_4 = F_2^2 \left[ \frac{\lambda_1}{2}\, r^2
  + \left| \psi \right| \epsilon
  \cos \left( \arg \psi + \arg I_{3/2} \right) r^{3/2}
  + \left( \lambda_3 + \frac{3}{4}\, \lambda_4 \delta \right) r
  + \frac{\lambda_2 + 2 \lambda_5 \gamma_5}{2}
  \right].
\label{pot1}
\ee

\paragraph{Bounds on $\epsilon^2$:}
In the gauge where $a = 0$ throughout space--time,
$I_{3/2} = b^3 f^\ast$
and therefore $\epsilon^2 F_1^3 F_2 = B^3 F = F_1^3 F$.
Hence,
\be
\epsilon^2 = \frac{F}{F + E + D + C}.
\label{17b}
\ee
Utilizing Eq.~\eqref{17b} together with Eq.~\eqref{deltaa=0} for $\delta$
in the same gauge,
one finds that
\bs
\ba
2 \epsilon^2 - 3 \delta + 1 &=& \frac{4 C + 2 D}{F + E + D + C},
\\
1 + \delta - 2 \epsilon^2 &=& \frac{2 D + E}{3 \left( F + E + D + C \right)}
\ea
\es
are both non-negative.
Therefore,
\be
3 \delta - 1 \le 2 \epsilon^2 \le 1 + \delta.
\label{analytical}
\ee
Thus,
the maximum possible value of $\epsilon$ is $1$
and it occurs only when $\delta = 1$.
Moreover,
$\delta = -1\ \Rightarrow\ \epsilon = 0$.

\paragraph{Parameters $x$, $y$, and $z$:}
In order to find the phase-space boundary,
it is expedient to assume the fields $c$,
$d$,
$e$,
and $f$ to be real\footnote{In all the cases studied in this paper,
we have worked numerically with complex fields
and found that the phase-space boundaries are the same for complex
and real fields.}
and to define
\bs
\label{xyz}
\ba
x &=& \frac{c}{\sqrt{c^2 + d^2 + e^2 + f^2}},
\\
y &=& \frac{d}{\sqrt{c^2 + d^2 + e^2 + f^2}},
\\
z &=& \frac{e}{\sqrt{c^2 + d^2 + e^2 + f^2}}.
\ea
\es
Then,
\be
x^2 + y^2 + z^2 = \frac{c^2 + d^2 + e^2}{c^2 + d^2 + e^2 + f^2} \le 1
\ee
and\footnote{Without loss of generality one may assume $f$ to be non-negative.}
\be
\frac{f}{\sqrt{c^2 + d^2 + e^2 + f^2}} = \sqrt{1 - x^2 - y^2 - z^2}.
\ee
Moreover,
\bs
\label{15}
\ba
\delta &=& 1 - 2 x^2 - \frac{4}{3}\, y^2 - \frac{2}{3}\, z^2,
\\
\gamma_5 &=&
\frac{1}{5} \left( 3 x \sqrt{1 - x^2 - y^2 - z^2} - y z \right)^2 \no & &
+ \frac{2}{5} \left\{ \left( \sqrt{3}\, x z - y^2 \right)^2
+ \left[ y\, \sqrt{3 \left( 1 - x^2 - y^2 - z^2 \right)}
  - z^2 \right]^2 \right\},
\ea
\es
and
\be
\label{pbvo}
\epsilon = \sqrt{1 - x^2 - y^2 - z^2}.
\ee

\paragraph{Boundary:}
Referring to Eqs.~\eqref{15} and~\eqref{pbvo},
if
\be
\label{mvfkk2}
\det \left( \begin{array}{ccc}
  \displaystyle{\frac{\partial \delta}{\partial x}} &
  \displaystyle{\frac{\partial \delta}{\partial y}} &
  \displaystyle{\frac{\partial \delta}{\partial z}} \\*[3mm]
  \displaystyle{\frac{\partial \gamma_5}{\partial x}} &
  \displaystyle{\frac{\partial \gamma_5}{\partial y}} &
  \displaystyle{\frac{\partial \gamma_5}{\partial z}} \\*[3mm]
  \displaystyle{\frac{\partial \epsilon}{\partial x}} &
  \displaystyle{\frac{\partial \epsilon}{\partial y}} &
  \displaystyle{\frac{\partial \epsilon}{\partial z}}
\end{array} \right) = 0,
\ee
then one is at the boundary
of the
$\left( \delta, \gamma_5, \epsilon \right)$
phase space.
Equation~\eqref{mvfkk2} states that the gradients of $\delta$,
  $\gamma_5$,
  and
  $\epsilon$
  are linearly dependent and thus span a surface---the border
  of phase space---instead of a volume,
  as they do when the determinant
  in the left-hand side of Eq.~\eqref{mvfkk2} is nonzero.
Equation~\eqref{mvfkk2} is an equation among $x$,
$y$,
and $z$ that one may convert into an equation among $\delta$,
$\gamma_5$,
and $\epsilon$ by inverting Eqs.~\eqref{15} and~\eqref{pbvo}.
In this way we have obtained the equation\footnote{Writing Eq.~\eqref{mvfkk2}
in terms of $\delta$,
$\gamma_5$,
and $\epsilon$ is a non-trivial task
that we were able to accomplish by having recourse
to the Gr\"obner basis.}
\be
\label{q2}
Q_1 \epsilon^2 Q_2 = 0,
\ee
where $Q_1$ is the quantity defined in Eq.~\eqref{14a}
and $Q_2$ is a polynomial function of $\delta$,
$\gamma_5$,
and $\epsilon^2$ explicitly given in Appendix~\ref{App:expressions}.
In practice,
the border of the allowed volume is formed by three sheets:
\begin{enumerate}
\item Sheet~1 has equation $Q_1 = 0$.
\item Sheet~2 has equation $\epsilon = 0$.
\item Sheet~3 has equation $Q_2 = 0$.
\end{enumerate}

\paragraph{Points:} We define the points
\bs
\ba
\widetilde P_1: & & \delta = -1,\  \gamma_5 = \epsilon = 0;
\\
\widetilde P_2: & & \delta = \frac{1}{3},\
\gamma_5 = \frac{2}{5},\ \epsilon = 0;
\\
\widetilde P_3: & & \delta = \epsilon = 1,\ \gamma_5 = 0;
\\
\widetilde P_4: & & \delta = \frac{3 + 4\, \sqrt{3}}{13},\
\gamma_5 = \frac{18 \left( 14 - 3\, \sqrt{3} \right)}{845},\
\epsilon^2 = \frac{11 + 6\, \sqrt{3}}{26};
\\
\widetilde P_5: & & \delta = 0,\
\gamma_5 = \frac{9}{20},\
\epsilon^2 = \frac{1}{2}.
\ea
\label{points_Y32}
\es
All these points are located on the boundary of phase space.
Both $\widetilde P_1$ and $\widetilde P_2$ have $Q_1 = Q_2 = \epsilon = 0$;
the other three points have $Q_1 = Q_2 = 0$ but $\epsilon \neq 0$.

\paragraph{Lines:} We define three lines linking $\widetilde P_1$
to $\widetilde P_2$ through different paths:
\begin{itemize}
\item The blue line has equation $Q_1 = \epsilon = 0$.
\item The purple line is given by $Q_2 = \epsilon = 0$.
  According to the expression for $Q_2$ given in Appendix~\ref{App:expressions},
  this means that for the purple line
  \bs
  \label{30}
  \ba
  256 \left( 10 \gamma_5 \right)^3
  - 32 \left( 89 + 90 \delta + 81 \delta^2 \right) \left( 10 \gamma_5 \right)^2
  & & \\
  + 9 \left( 1\,129 + 2\,868 \delta + 4\,614 \delta^2
  + 3\,348 \delta^3  + 729 \delta^4 \right) \left( 10 \gamma_5 \right)
  & & \\
  - 81 \left( 1 + \delta \right)^3
  \left( 127 + 225 \delta + 405 \delta^2 + 243 \delta^3 \right) &=& 0.
  \ea
  \es
  Equation~\eqref{30} implicitly gives,
  together with $\epsilon = 0$,
  the purple line.
\item The magenta--green--brown line is given by $Q_1 = Q_2 = 0$.
  It has three segments:
  \begin{itemize}
  \item The magenta segment begins at $\widetilde P_1$
    and ends at $\widetilde P_3$.
    It has
    \be
    \epsilon^2 = \frac{1 + \delta}{2}.
    \ee
    The point
    $\widetilde P_5$
    lies between $\widetilde P_1$ and $\widetilde P_3$ on the magenta segment.
  \item The green segment begins at $\widetilde P_3$
    and ends at $\widetilde P_4$.
    It has
    \be
    \epsilon^2 = \frac{1 + 3 \delta}{4}.
    \ee
  \item The brown segment begins at $\widetilde P_4$
    and ends at $\widetilde P_2$.
    It has
    \be
    \epsilon^2 = \frac{\left( 3 \delta - 1 \right)
      \left( 7 \delta + 3 \right)^3}{2\,048\, \delta^3}.
    \ee
  \end{itemize}
  All three segments have $Q_1 = 0$,
  \textit{viz.}\ $\gamma_5 = \left. 9 \left( 1 - \delta^2 \right) \right/ 20$.
  We note that at their meeting point $\widetilde P_4$
  the green and brown segments have continuous first derivative
  \be
  \left. \frac{\mathrm{d} \epsilon}{\mathrm{d} \delta} \right|_{\widetilde P_4}
  = \frac{3 \sqrt{2}}{8}\, \sqrt{11 - 6\sqrt{3}},
  \ee
  but different second derivatives:
  \bs
  \ba
  \left. \frac{\mathrm{d}^2 \epsilon_{\mathrm{green}}}{\left(\mathrm{d}
    \delta\right)^2} \right|_{\widetilde P_4} &=& -\frac{117 \sqrt{26}}{32
    \left(\sqrt{11 + 6\sqrt{3}}\right)^3},
  \\
  \left. \frac{\mathrm{d}^2 \epsilon_{\mathrm{brown}}}{\left(\mathrm{d}
    \delta\right)^2} \right|_{\widetilde P_4}
  &=& \frac{2 \left(9 + 7\sqrt{3}\right)}{9}\,
  \left. \frac{\mathrm{d}^2 \epsilon_{\mathrm{green}}}{\left(\mathrm{d}
    \delta\right)^2} \right|_{\widetilde P_4}.
  \ea
  \es
  On the other hand,
  at the point $\widetilde P_3$ the magenta--green--brown line changes
  $\mathrm{d}
  \epsilon
  \left/ \mathrm{d} \delta \right.$.
  Referring to the inequalities~\eqref{analytical}
  we note that the upper bound $2 \epsilon^2 \le 1 + \delta$
  is saturated at the magenta segment of this line.
\end{itemize}

\paragraph{Bounding surface:} The surface bounding phase space
has the topology of the surface of a sphere.
The ``poles'' are $\widetilde P_1$ and $\widetilde P_2$.
The surface is formed by sheet~1,
followed by the blue line,
followed by sheet~2,
followed by the purple line,
followed by the sheet~3,
followed by the magenta--green--brown line,
then returning to sheet~1.

\paragraph{Auxiliary lines:}
It is useful to define two other lines,
which have $Q_2 = 0$ but neither $Q_1 = 0$ nor $\epsilon = 0$,
\textit{i.e.}\ they lie on sheet~3.
\begin{itemize}
\item The orange line is defined by
  \bs
  \ba
  \gamma_5 &=& \frac{9 \left( \delta - 1 \right)^2}{10},
  \\
  \epsilon^2 &=& \frac{3 \delta - 1}{2}.
  \ea
  \es
  The orange line starts at $\widetilde P_2$ and ends at $\widetilde P_3$.
  The lower bound $2 \epsilon^2 \ge 3 \delta - 1$
  is saturated at the orange line for $1/3 \le \delta \le 1$;
  for $-1 \le \delta \le 1/3$
  the lower bound on $\epsilon^2$ is simply $\epsilon^2 \ge 0$
  and is saturated at sheet~2.
\item The yellow line is defined by
  \bs
  \ba
  \gamma_5 &=& 0,
  \\
  \epsilon^2 &=& \frac{\left( 1 + \delta \right)^3}{8}.
  \ea
  \es
  The yellow line starts at $\widetilde P_1$ and ends at $\widetilde P_3$.
  It saturates the bound $\gamma_5 \ge 0$.
\end{itemize}

\begin{figure}[h!]
\begin{center}
\includegraphics[width=0.45\textwidth]{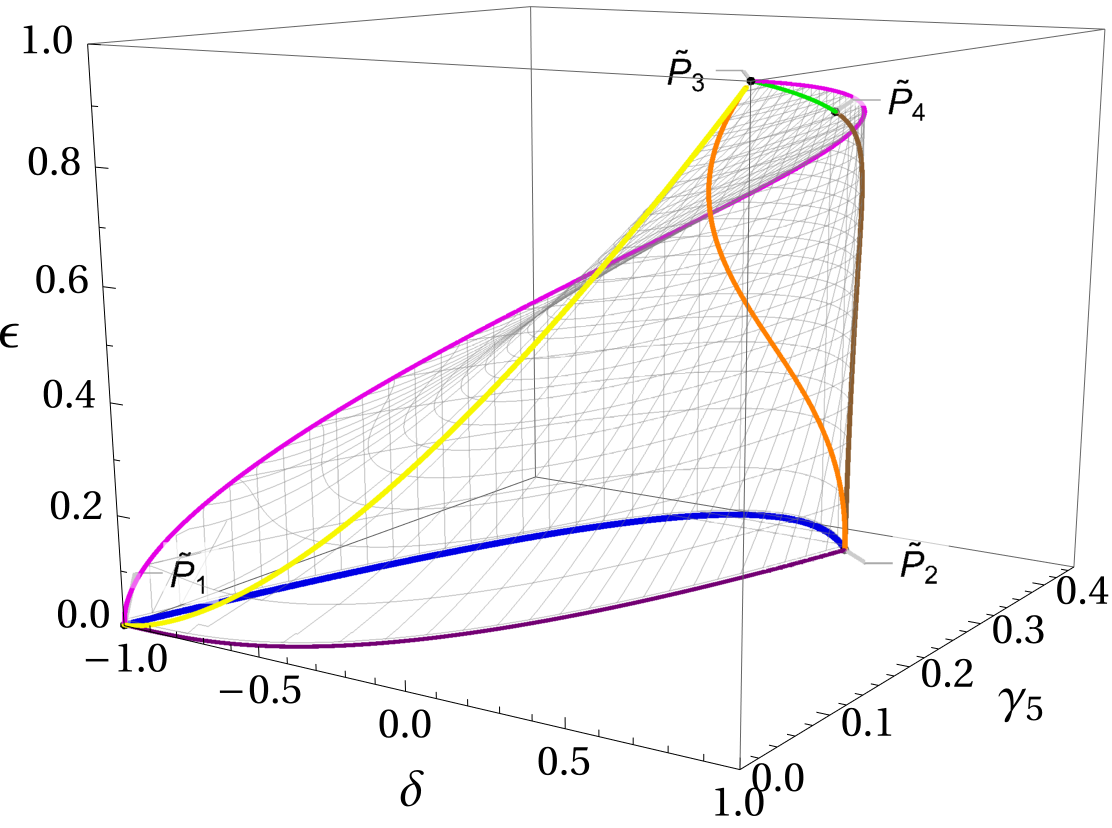}
\hspace{0.05\textwidth}
\includegraphics[width=0.465\textwidth]{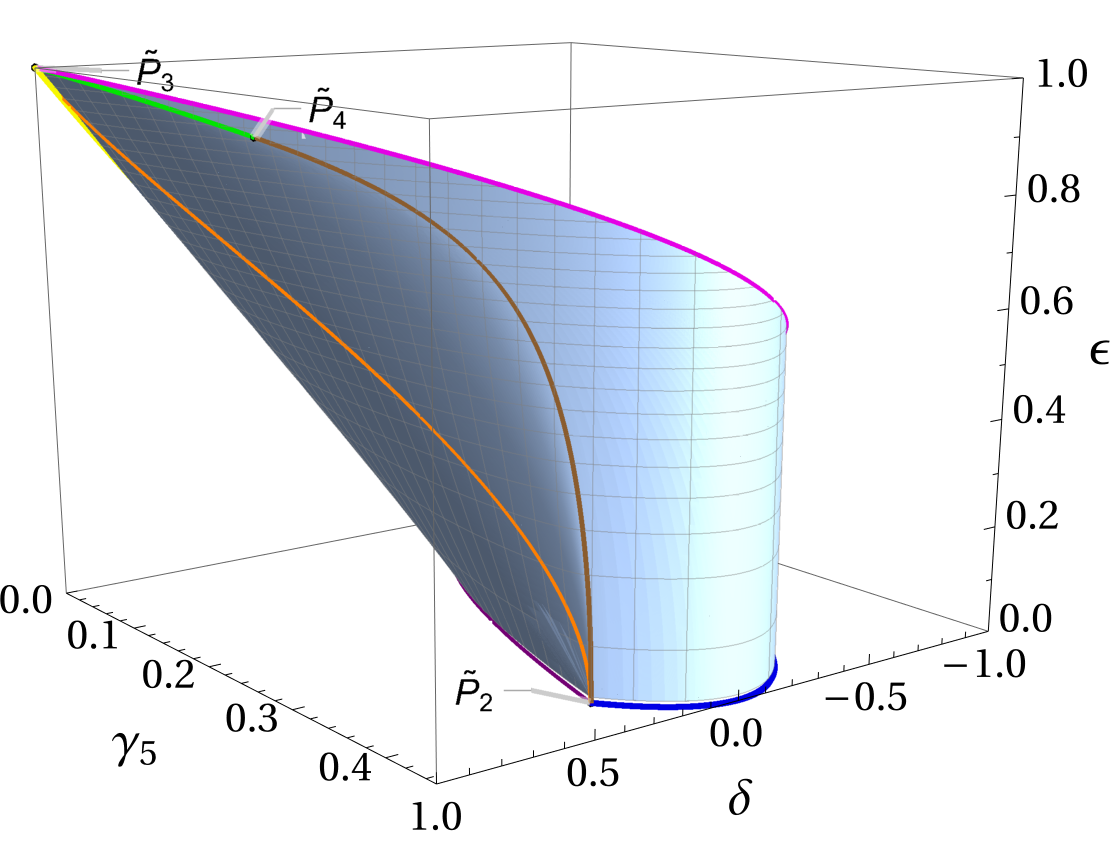}
\end{center}
\caption{Two perspectives of the phase space for case $Y=3/2$.
  The lines and the points---except
  $\widetilde P_5$---defined in the text
  are explicitly displayed.
  In the left panel the phase space is displayed transparent,
  so that the lines may be seen behind each other;
  in the right panel it is opaque.}
\label{fig:Y32-3D}
\end{figure}

\begin{figure}[h!]
\begin{center}
\includegraphics[width=0.45\textwidth]{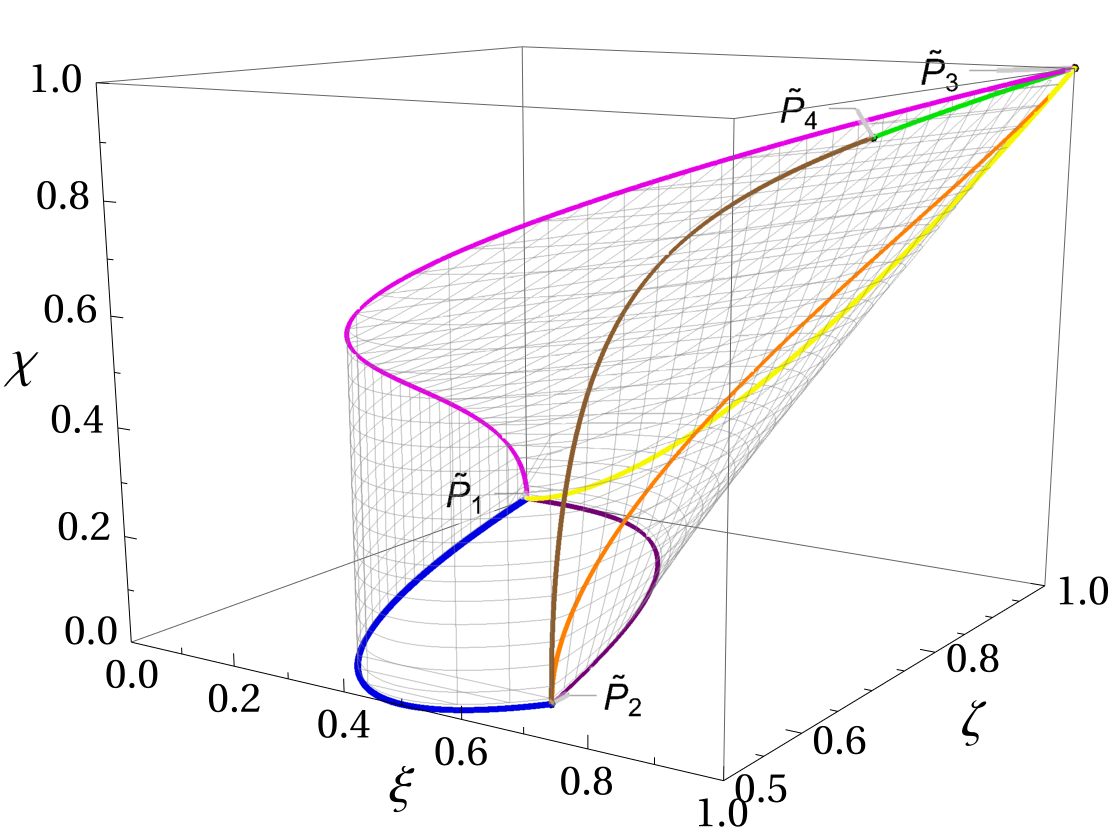}
\hspace{0.05\textwidth}
\includegraphics[width=0.465\textwidth]{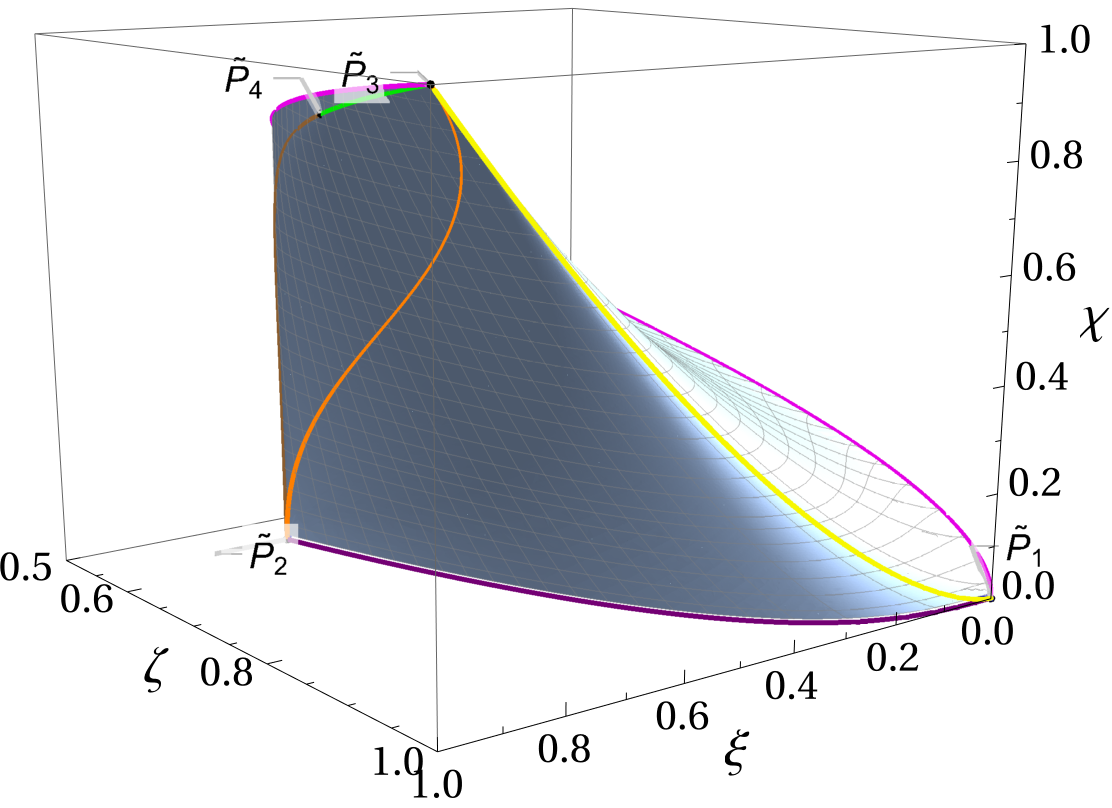}
\end{center}
\caption{The same as Fig.~\ref{fig:Y32-3D}
      but using the parameters $\xi$, $\zeta$, and $\chi$ of Ref.~\cite{kannike}
      instead of our parameters $\delta$, $\gamma_5$, and $\epsilon$.
      Note that $\chi$ coincides with our parameter $\epsilon$.
      \vspace{5mm}}
\label{fig:xi-zeta}
\end{figure}

\begin{figure}[h!]
\begin{center}
\includegraphics[width=1.0\textwidth]{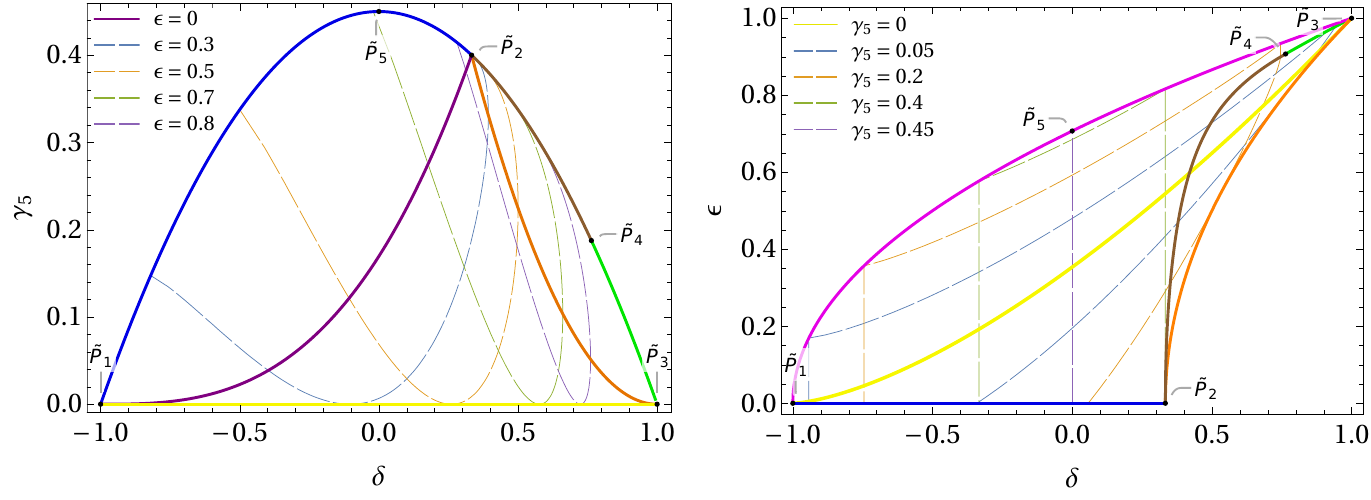}
\end{center}
\caption{The projections of the phase space for case $Y=3/2$
  on the planes $\gamma_5$ \textit{vs.}\ $\delta$ (left panel)
  and
  $\epsilon$
  \textit{vs.}\ $\delta$ (right panel).
  See further comments on this figure in the text.}
\label{fig:Y32-proj1}
\end{figure}

\paragraph{Figures:} Figure~\ref{fig:Y32-3D} has two panels
with different perspectives of phase space.
Figure~\ref{fig:xi-zeta} is the same as Fig.~\ref{fig:Y32-3D}
but using the parameters $\xi$ and $\zeta$ of Ref.~\cite{kannike}
instead of our parameters $\delta$ and $\gamma_5$,
\textit{cf.}\ Eqs.~\eqref{zeta}.
Figure~\ref{fig:Y32-proj1} shows two projections of the phase space on planes.
On the left panel of that figure,
note the following.
\begin{enumerate}
\item The magenta segment coincides with the union of the blue line,
  the brown segment,
  and the green segment.
\item Sheet~1 is wholly projected on the blue line
  and on the brown and green segments.
  Sheet~2 is the left part of the surface,
  between the blue and purple lines.
  Sheet~3 is the whole space under the parabola.
\item The dashed lines
  (and the full purple line for
  $\epsilon = 0$)
  display the contours of constant
  $\epsilon$
  \emph{on sheet~3}.
\end{enumerate}
On the right panel of Fig.~\ref{fig:Y32-proj1} note the following.
\begin{enumerate}
\item The purple line coincides with the blue line.
\item Sheet~2 is wholly projected onto the blue line.
  Sheet~3 occupies the whole triangle among the blue and orange lines
  and the magenta segment.
  Sheet~1 only occupies the left part of the triangle,
  between the blue line and the magenta--green--brown line.
\item Besides the full yellow line with $\gamma_5 = 0$,
  there are dashed lines of constant $\gamma_5$;
  they are vertical on sheet~1 and curved on sheet~3,
  explicitly displaying the concavity of that sheet.
\end{enumerate}

\paragraph{Boundedness from below:}
The potential is BFB if and only if $V_4 \ge 0$
for any values of the fields.\footnote{A stricter definition
of boundedness-from-below requires $V_4$ to be strictly positive.
In this paper we are a bit sloppy;
we do not pay much attention
to the possibilities of strict equality,
which are anyway meaningless under renormalization.}
Referring to Eq.~\eqref{pot1} and taking into account that the fields
may take phases such that $\cos \left( \arg \psi  + \arg I_{3/2} \right) = -1$,
we require that
\be
\frac{V_4}{F_2^2} =
c_4 \left( \sqrt{r} \right)^4
+ c_3 \left( \sqrt{r} \right)^3
+ c_2 \left( \sqrt{r} \right)^2
+ c_0 \ge 0
\label{pot25}
\ee
for every $\sqrt{r} > 0$,
where
\be
c_4 = \frac{\lambda_1}{2}, \quad
c_3 = - \left| \psi \right| \epsilon, \quad
c_2 = \lambda_3 + \frac{3}{4}\, \lambda_4 \delta, \quad
c_0 = \frac{\lambda_2 + 2 \lambda_5 \gamma_5}{2}.
\label{pot26}
\ee
Note the crucial facts that there is no term $c_1 \sqrt{r}$
in Eq.~\eqref{pot25} and that $c_3 \le 0$ in Eq.~\eqref{pot26}.
By considering the cases $\sqrt{r} \to \infty$ and $\sqrt{r} \to 0$,
condition~\eqref{pot25} of course requires $c_4 \ge 0$ and $c_0 \ge 0$,
respectively,
\textit{i.e.}
\bs
\label{29}
\ba
\lambda_1 &\ge& 0,
\label{29a} \\
\lambda_2 + 2 \lambda_5 \gamma_5 &\ge& 0.
\label{29b}
\ea
\es
Since the left-hand side of condition~\eqref{29b}
is a monotonic function of $\gamma_5$,
that condition holds for all $\gamma_5$ if and only if it holds for the
maximal and minimal $\gamma_5$,
\textit{i.e.}\ if conditions
\bs
\label{300}
\ba
\lambda_2 &\ge& 0,
\\
\lambda_2 + \frac{9}{10}\, \lambda_5 &\ge& 0
\ea
\es
hold true.
We next apply Theorem~2 of Ref.~\cite{quartic},
which states\footnote{The theorem also applies to the case $c_1 \neq 0$,
so we are using here a simplified version thereof.}
that condition~\eqref{pot25} holds $\forall \sqrt{r} > 0$ if and only if
\begin{enumerate}
\item either
  \be
  \Delta \le 0\ \mbox{and}\ \aleph > 0,
  \label{uocppv}
  \ee
\item or
  \be
  \Delta \ge 0\ \mbox{and}\
  \left\{ \begin{array}{l}
      \mbox{either}\ -2 \le \beth \le 6\ \mbox{and}\ \Lambda_1 \le 0,
      \\*[1mm]
      \mbox{or}\ \beth > 6\ \mbox{and}\ \Lambda_2 \le 0.
  \end{array} \right.
  \label{uocppv2}
  \ee
\end{enumerate}
Here,
\bs
\ba
\Delta &=& 4 \left( \beth^2 + 12 \right)^3
- \left( 2 \beth^3 - 72 \beth + 27 \aleph^2 \right)^2,
\label{eq:Delta}
\\
\Lambda_1 &=& \aleph^2 - 16 \aleph - 16 \left( \beth + 2 \right),
\label{eq:Lambda1}
\\
\Lambda_2 &=& \aleph^2 - 4 \aleph\ \frac{\beth + 2}{\sqrt{\beth - 2}}
- 16 \left( \beth + 2 \right),
\label{eq:Lambda2}
\ea
\label{eq:Delta_Lambdas}
\es
where
\bs
\label{aleph-beth}
\ba
\aleph &=& \frac{c_3}{\sqrt[4]{c_4^3 c_0}},
\\
\beth &=& \frac{c_2}{\sqrt{c_4 c_0}}.
\ea
\es
Since $c_3 \le 0$,
$\aleph \le 0$ and possibility~\eqref{uocppv} may be discarded,
so only option~\eqref{uocppv2} remains.
That option necessitates $\beth \ge -2$,
which is precisely the condition $c_2 \ge - 2 \sqrt{c_4 c_0}$
that also applies when $c_3 = 0$.
Hence,
the following necessary BFB conditions~\cite{milagre} apply:
\bs
\label{fj9d99}
\ba
\lambda_3 - \frac{3 \left| \lambda_4 \right|}{4} + \sqrt{\lambda_1 \lambda_2}
&\ge& 0,
\\
\lambda_3 + \sqrt{\lambda_1 \left( \lambda_2 + \frac{9}{10}\, \lambda_5 \right)}
&\ge& 0.
\ea
\es
Furthermore,
one must \emph{impede} the situation where
\bs
\label{imped}
\ba
8 \lambda_1 \lambda_5 &<& - 5 \lambda_4^2,
\\
6 \sqrt{\lambda_1}\, \lambda_5 &<&
- 5 \sqrt{\lambda_2} \left| \lambda_4 \right|,
\\
\lambda_3 &<& - \sqrt{\frac{\left( 10 \lambda_2 + 9 \lambda_5 \right)
    \left( 8 \lambda_1 \lambda_5 + 5 \lambda_4^2 \right)}{80 \lambda_5}}
\ea
\es
\emph{simultaneously}.
We next consider the constraints on $\aleph$ in condition~\eqref{uocppv2}.
Since $\Delta \ge 0$,
\be
\frac{2}{27} \left[ \beth \left( 36 - \beth^2 \right)
  - \left( \sqrt{\beth^2 + 12} \right)^3 \right]
\le \aleph^2 \le
\frac{2}{27} \left[ \beth \left( 36 - \beth^2 \right)
  + \left( \sqrt{\beth^2 + 12} \right)^3 \right].
\label{hj22}
\ee
Furthermore,
either
\be
-2 \le \beth \le 6 \quad \mbox{and} \quad
0 \ge \aleph \ge - 4 \left( \sqrt{\beth + 6} - 2 \right)
\label{hj23}
\ee
or
\be
\beth > 6 \quad \mbox{and} \quad
0 \ge \aleph \ge - 2\ \sqrt{\frac{\beth + 2}{\beth - 2}}\,
\left( \sqrt{5 \beth - 6} - \sqrt{\beth + 2} \right).
\label{hj24}
\ee
Since inequalities~\eqref{hj23} and~\eqref{hj24}
effectively impose lower bounds on $\aleph$,
they must apply to
\be
\aleph = \frac{ - 2 \left| \psi \right|
  \epsilon_\mathrm{max} \left( \delta, \gamma_5 \right)}{\sqrt[4]{\lambda_1^3
    \left( \lambda_2 + 2 \lambda_5 \gamma_5 \right)}},
\label{aleph1}
\ee
where $\epsilon_\mathrm{max} \left( \delta, \gamma_5 \right)$
is the maximum possible value of $\epsilon$ for each $\delta$ and $\gamma_5$;
clearly,
this is always given by a solution of equation $Q_2 = 0$,
hence the exact expression for $Q_2$ in Appendix~\ref{App:expressions}
is relevant.

\paragraph{Importance of the magenta line:} 
By following the procedure described in Appendix~\ref{App:concavity}
we have found that
sheet~3
has both a (rather small)
concave part and a (much larger) part which is neither concave nor convex;
sheet~3
has no convex part. 
As a consequence,
one needs to check conditions~\eqref{hj22}--\eqref{hj24} 
\emph{only along the boundaries of the surface},
specifically along the magenta and yellow lines.
As a matter of fact,
our numerical computations---performed with $10^6$ randomly generated
sets of parameters
$\lambda_1, \ldots, \lambda_5, \left| \psi \right|$---indicate that
the scan along the yellow line is superfluous:
it is sufficient to scan \emph{the magenta line},
\textit{i.e.}\
it is sufficient to use
\bs
\label{alephss}
\ba
\aleph &=& \frac{- \left| \psi \right|
  \sqrt{2 \left( 1 + \delta \right)}}{\sqrt[4]{\lambda_1^3 \left[ \lambda_2
      + \left. 9 \left( 1 - \delta^2 \right) \lambda_5 \right/ 10 \right]}},
\\
\beth &=& \frac{4 \lambda_3 + 3 \lambda_4 \delta}{2
  \sqrt{\lambda_1 \left[ \lambda_2 + \left. 9
      \left( 1 - \delta^2 \right) \lambda_5 \right/ 10 \right]}}
\ea
\es
with $-1 \le \delta \le 1$.
Thus,
our practical recipe for ascertaining the boundedness-from-below of $V_4$
consists of the following successive steps:
\begin{enumerate}
\item Checking the necessary conditions~\eqref{29a},
  \eqref{300},
  and~\eqref{fj9d99}.
\item Discarding any case where conditions~\eqref{imped} apply.
\item Testing conditions~\eqref{hj22}--\eqref{hj24}
  at the special points~\eqref{points_Y32}.
\item Testing conditions~\eqref{hj22}--\eqref{hj24}
  with $\aleph$ and $\beth$ given by Eqs.~\eqref{alephss}
  for $\delta \in \left[-1, 1 \right]$.
\end{enumerate}
Step~4 must, of course, be performed numerically
and with sufficiently small steps of $\delta$.

\paragraph{Efficiency of the method:}
In order to demonstrate the efficiency of our method,
we have compared it with the direct minimization of
$V_4$.
We have randomly generated $10^6$ sets of couplings
$\lambda_3, \lambda_4, \lambda_5, \psi \in \left[ -10, 10 \right]$
and $\lambda_1, \lambda_2 \in \left[ 0, 10 \right]$,
being careful to ensure that
the distribution of each coupling over its respective range is uniform.
The potential $V_4$ was then minimized\footnote{In order to ensure
high accuracy,
we have employed two minimization algorithms---specifically,
the Nelder--Mead method
and the Differential Evolution method.
The minimization was made over \emph{real} fields $a, b, \ldots, f$,
letting those fields vary over various ranges---for instance,
from $-1$ to $1$,
from $-100 $ to $100$,
and from $-10^5$ to $10^5$;
by doing this we aimed at not missing any minimum,
independently of it having small or large values of the fields.}
for each set of couplings.
Alternatively,
we have used
our recipe given in the previous paragraph.
Our results were:
\begin{itemize}
\item The minimization of $V_4$ for the $10^6$ sets of parameters
  required 39\,516 seconds,\footnote{Our computations
  were performed on a desktop computer
  equipped with an Intel Core i9-13900K CPU featuring 24 cores.
  All the cases being compared were evaluated under identical conditions
  by using {\tt Mathematica},
  with calculations executed in parallel across all cores.}
  producing 256\,179 quartic potentials that are bounded from below.
\item Our scanning method completed the task in 29.2 seconds,
  yielding exactly the same 256\,179 BFB potentials.
\end{itemize}
This comparison demonstrates that our method is approximately 1\,300 times
faster than the brute-force minimization of the potential
and suggests that it is just as exact.

\section{Quadruplet with hypercharge $1/2$ and a reflection symmetry}
\label{sec:1/2}

\paragraph{Extra term:}
Multiplying the doublet $\Phi$ by itself
one obtains the $SU(2)$ triplet
\be
T_3 = \left( \Phi \otimes \Phi \right)_\mathbf{3}
= \left( \begin{array}{c} a^2 \\ \sqrt{2}\, a b \\ b^2 \end{array} \right).
\label{tir3}
\ee
Multiplying the quadruplet $\Xi$ by itself one also obtains
(besides a seven-plet) the $SU(2)$ triplet $T_1$ of Eq.~\eqref{tir1}.
%%
%\be
%\bar T_2 = \left( \Xi \otimes \Xi \right)_\mathbf{3} =
%\frac{1}{\sqrt{5}} \left( \begin{array}{c}
%  \sqrt{6}\, c e - \sqrt{2}\, d^2 \\
%  3 c f - d e \\
%  \sqrt{6}\, d f - \sqrt{2}\, e^2
%\end{array} \right).
%\label{ode9}
%\ee
%h
If the hypercharge of $\Xi$ is the same as the one of $\Phi$,
\textit{i.e.}\ if $\Xi$ has hypercharge $1/2$,
then one may construct an $SU(2) \times U(1)$-invariant quantity
out of the triplets~\eqref{tir1} and~\eqref{tir3}:
\be
I_{1/2} = \sqrt{\frac{2}{5}} \left[
  {a^\ast}^2 \left( \sqrt{3}\, c e - d^2 \right)
  + a^\ast b^\ast \left( 3 c f - d e \right)
  + {b^\ast}^2 \left( \sqrt{3}\, d f - e^2 \right) \right].
\ee
If moreover the Lagrangian is symmetric
under $\Xi \to - \Xi$,\footnote{The extra symmetry is needed
to remove the terms
\bs
\label{gi9f0f0}
\ba
& & \left[ \left( \Xi \otimes \Xi \right)_\mathbf{3} \otimes
  \left( \tilde \Xi \otimes \tilde \Phi \right)_\mathbf{3}
  \right]_\mathbf{1},
\\
& & \left[ \left( \Phi \otimes \Phi \right)_\mathbf{3} \otimes
  \left( \tilde \Xi \otimes \tilde \Phi \right)_\mathbf{3}
  \right]_\mathbf{1},
\ea
\es
that in general are also present in the SP
if $\Xi$ and $\Phi$ have the same hypercharge.
In the expressions~\eqref{gi9f0f0},
$\tilde \Phi$ and $\tilde \Xi$
are the $SU(2)$ doublet and the $SU(2)$ quadruplet,
respectively,
formed by the complex conjugates of the fields $a, \ldots, f$.}
then
\be
V_{4,\mathrm{extra}} = \frac{\chi}{2}\, I_{1/2} + \mathrm{H.c.},
\label{03kkfd}
\ee
where $\chi$ is a complex dimensionless coefficient.
We define the dimensionless non-negative real parameter $\eta$ as
\be
\eta = \frac{\left| I_{1/2} \right|}{F_1 F_2}.
\ee
Then,
\be
V_{4,\mathrm{extra}} =
\left| \chi \right| \eta F_1 F_2 \cos \left( \arg \chi + \arg I_{1/2} \right),
\ee
so that
\be
V_4 = F_2^2 \left\{ \frac{\lambda_1}{2}\, r^2
+ \left[ \lambda_3 + \frac{3}{4}\, \lambda_4 \delta
  + \left| \chi \right| \eta \cos \left( \arg \chi + \arg I_{1/2} \right)
  \right] r
+ \frac{\lambda_2 + 2 \lambda_5 \gamma_5}{2}
\right\}.
\label{pot2}
\ee

\paragraph{Upper bounds on $\eta$:}
In the gauge where $a=0$,
$I_{1/2} = \sqrt{2/5}\ {b^\ast}^2 \left( \sqrt{3}\, d f - e^2 \right)$
and therefore $\eta^2 F_1^2 F_2^2
= \left( 2/5 \right) B^2 \left| \sqrt{3}\, d f - e^2 \right|^2
= \left( 2/5 \right) F_1^2 \left| \sqrt{3}\, d f - e^2 \right|^2$.
Hence,
\be
\eta^2 = \frac{2 \left| \sqrt{3}\, d f - e^2 \right|^2}{5 \left(
  F + E + D + C \right)^2},
\label{17bbb}
\ee
which should be used together with Eq.~\eqref{deltaa=0}.
One then finds that
\bs
\ba
Q_3 &=& 9 \left( 1 + \delta \right)^2 - 40 \eta^2
\label{q3}
\\ &=& \frac{4 \left( 3 F - D \right)^2 + 16 \left| d^\ast e + \sqrt{3}\,
  e^\ast f \right|^2}{\left( F + E + D + C \right)^2}
\\
&\ge& 0.
\ea
\es
Therefore,
\be
\eta^2 \le \frac{9 \left( 1 + \delta \right)^2}{40}.
\label{upper1}
\ee
Another upper bound on $\eta$ is
\be
\eta^2 \le \gamma_5,
\label{upper2}
\ee
which directly follows from Eqs.~\eqref{f2},
\eqref{f5},
\eqref{g5},
and~\eqref{17bbb}.

\paragraph{Boundary:}
We assume $c$,
$d$,
$e$,
and $f$ to be real instead of complex and we define the real parameters $x$,
$y$,
and $z$ through Eqs.~\eqref{xyz}.
Then,
\be
\label{mvfkk3}
\eta^2 = \frac{2}{5} \left[ y\, \sqrt{3 \left( 1 - x^2 - y^2 - z^2 \right)}
  - z^2 \right]^2,
\ee
which should be used together with Eqs.~\eqref{15}.
A point where
\be
\label{mvfkk5}
\det \left( \begin{array}{ccc}
  \displaystyle{\frac{\partial \delta}{\partial x}} &
  \displaystyle{\frac{\partial \delta}{\partial y}} &
  \displaystyle{\frac{\partial \delta}{\partial z}} \\*[3mm]
  \displaystyle{\frac{\partial \gamma_5}{\partial x}} &
  \displaystyle{\frac{\partial \gamma_5}{\partial y}} &
  \displaystyle{\frac{\partial \gamma_5}{\partial z}} \\*[3mm]
  \displaystyle{\frac{\partial \eta^2}{\partial x}} &
  \displaystyle{\frac{\partial \eta^2}{\partial y}} &
  \displaystyle{\frac{\partial \eta^2}{\partial z}}
\end{array} \right) = 0
\ee
is a candidate to lie at the boundary of the
$\left( \delta, \gamma_5, \eta \right)$ phase space.
Equation~\eqref{mvfkk5} translates into
\be
Q_1 Q_3 Q_4 \eta^2 = 0,
\label{jvfdodo}
\ee
where $Q_1$ was defined in Eq.~\eqref{14a},
$Q_3$ was defined in Eq.~\eqref{q3},
and $Q_4$ is a polynomial function of $\delta$,
$\gamma_5$,
and $\eta^2$ that is explicitly written down in Appendix~\ref{App:expressions}.
In practice,
the boundary of phase space is formed by four sheets:
\begin{enumerate}
\item Sheet~1 has equation $Q_1 = 0$.
\item Sheet~2 has equation $Q_3 = 0$.
\item Sheet~3 has equation $Q_4 = 0$.
\item Sheet~4 has equation $\eta = 0$.
\end{enumerate}

\paragraph{Points:} We define the points
\bs
\label{points_Y12}
\ba
\overline P_1: & & \delta = -1,\ \gamma_5 = \eta = 0;
\\
\overline P_2: & & \delta = \frac{1}{3},\ \gamma_5  = \eta^2 = \frac{2}{5};
\\
\overline P_3: & & \delta = \frac{1 + 2 \sqrt{2}}{7},\
\gamma_5 = \frac{9 \left( 10 - \sqrt{2} \right)}{245},\
\eta^2 = \frac{9 \left( 9 + 4 \sqrt{2} \right)}{490};
\\
\overline P_4: & & \delta = \frac{2}{3},\ \gamma_5 = \frac{1}{4},\
\eta^2 = \frac{9}{40};
\\
\overline P_5: & & \delta= 1,\ \gamma_5 = \eta = 0.
\ea
\es
All these points are located on the boundary of phase space.
Point $\overline P_1$ has $Q_1 = Q_3 = Q_4 = \eta = 0$.
Point $\overline P_2$ has $Q_1 = Q_3 = Q_4 = 0$ but $\eta \neq 0$.
Point $\overline P_5$ has $Q_1 = Q_4 = \eta = 0$ but $Q_3 \neq 0$.
Both point $\overline P_3$ and point $\overline P_4$
have $Q_1 = Q_4 = 0$ but $Q_3 \neq 0$ and $\eta \neq 0$.

\paragraph{Lines:}
We define two lines linking $\overline P_1$ to $\overline P_2$
through different paths:
\begin{itemize}
\item The magenta line is given by $Q_1 = Q_3 = 0$.
\item The purple line is given by $Q_3 = Q_4 = 0$,
  \textit{viz.}\ by
  \bs
  \ba
  \gamma_5 &=&
  \frac{9 \left( 3 - \delta \right) \left( 1 + \delta \right)}{80},
  \\
  \eta &=& \frac{3 \left( 1 + \delta \right)}{2 \sqrt{10}}.
  \ea
  \es
\end{itemize}
We furthermore define two lines linking $\overline P_1$ to $\overline P_5$
through different paths:
\begin{itemize}
\item The blue line is given by $Q_1 = \eta = 0$.
\item The yellow line is given by $Q_4 = \eta = \gamma_5 = 0$.
\end{itemize}
Finally,
we define the green--brown line,
which has $Q_1 = Q_4 = 0$ and connects $\overline P_2$ to $\overline P_5$.
This line has two segments\footnote{The equations for the green segment
and for the brown segment are distinct solutions of $Q_1 = Q_4 = 0$.
Since $Q_4$ is a high-order polynomial,
it has various roots,
out of which two are relevant for sheet~3.}:
\begin{itemize}
\item The brown segment goes from $\overline P_2$ to $\overline P_3$.
  It has $Q_1 = 0$ and
  \be
  \eta = \frac{3 \left( 1 + \delta \right)^2}{8 \sqrt{10}\, \delta}.
  \ee
\item The green segment goes from $\overline P_3$ to $\overline P_5$.
  It has $Q_1 = 0$ and
  \be
  \eta = \frac{3
    \sqrt{ \left( 3 \delta + 1 \right) \left( 1 - \delta \right)}}{2 \sqrt{10}}.
  \ee
  The point $\overline P_4$ lies on the green segment.
\end{itemize}
At the point $\overline P_3$ where they meet,
the green and brown segments have equal first derivative
\be
\left. \frac{\mathrm{d} \eta}{\mathrm{d} \delta} \right|_{\overline P_3}
= - \frac{3 \sqrt{5}}{10}\sqrt{3 - 2\sqrt{2}},
\ee
but symmetric second derivatives:
\be
\left. \frac{\mathrm{d}^2 \eta_{\mathrm{brown}}}{\left(\mathrm{d}
  \delta\right)^2} \right|_{\overline P_3}
= - \left. \frac{\mathrm{d}^2 \eta_{\mathrm{green}}}{\left(\mathrm{d}
  \delta\right)^2} \right|_{\overline P_3}
= \frac{3 \sqrt{5}}{40}\, \sqrt{44 - 25\sqrt{2}}.
\ee

\paragraph{Bounding surface:} The surface that bounds phase space
has the topology of the surface of a sphere.
The three points $\overline P_1$,
$\overline P_2$,
and $\overline P_5$ form a triangle on that surface.
The sides of the triangle are the purple,
yellow,
and green--brown lines.
The magenta line and the blue line lie inside the triangle.
The whole surface outside the triangle has $Q_4 = 0$,
\textit{i.e.}\ it is sheet~3.
Inside the triangle there is sheet~2 between the purple and magenta lines,
sheet~4 between the yellow and blue lines,
and sheet~1 among the magenta,
blue,
and green--brown lines.

\paragraph{Orange line:}
The upper bound~\eqref{upper1} on $\eta^2$
gets saturated at the magenta line for $-1 \le \delta \le 1/3$.
For $1/3 \le \delta \le 1$ the relevant upper bound on $\eta^2$
is~\eqref{upper2},
which is saturated at the orange line.
This is defined by
\be
\gamma_5 = \eta^2 = \frac{9 \left( 1 - \delta \right)^2}{10}.
\ee
The orange line goes from $\overline P_2$ to $\overline P_5$
just as the green--brown line.
It has $Q_4 = 0$ but $Q_1 \neq 0$,
$Q_3 \neq 0$,
and $\eta \neq 0$,
so it is outside the triangle defined in the previous paragraph.

\paragraph{Concavity and the red line:}
Sheet~2 is flat:
it is on the plane
$2 \sqrt{10}\, \eta = 3 \left( 1 + \delta \right)$.
Following the procedure described in Appendix~\ref{App:concavity},
we have found that sheet~3 consists both of a convex region
and of a region that is neither concave nor convex.
These two regions are separated by the red line,
defined by
\bs
\label{red_line}
\ba
\gamma_5 &=& \frac{5 - \sqrt{7 - 18 \delta}}{10},
\\
\eta &=& \frac{8 + 3 \delta - \sqrt{7 - 18 \delta}}{4 \sqrt{10}}
\ea
\es
for $-1 \le \delta \le 1/3$.
The red line lies on the plane
$4 \sqrt{10}\, \eta = 3 + 3 \delta + 10 \gamma_5$.
The portion of sheet~3 between the purple and red lines
is neither convex nor concave;
the portion of sheet~3 between the red and green--brown lines is convex.

\begin{figure}[h!]
  \begin{center}
    \includegraphics[width=0.45\textwidth]{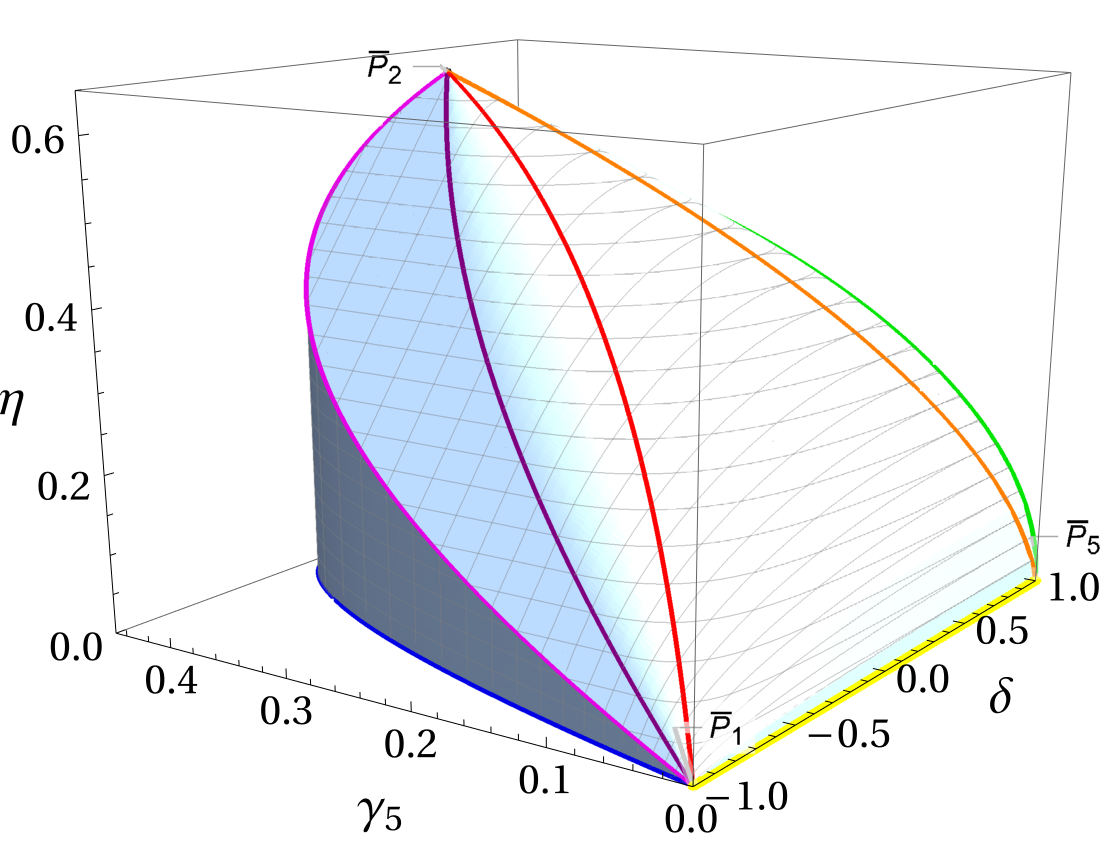}
    \hspace{0.05\textwidth}
    \includegraphics[width=0.45\textwidth]{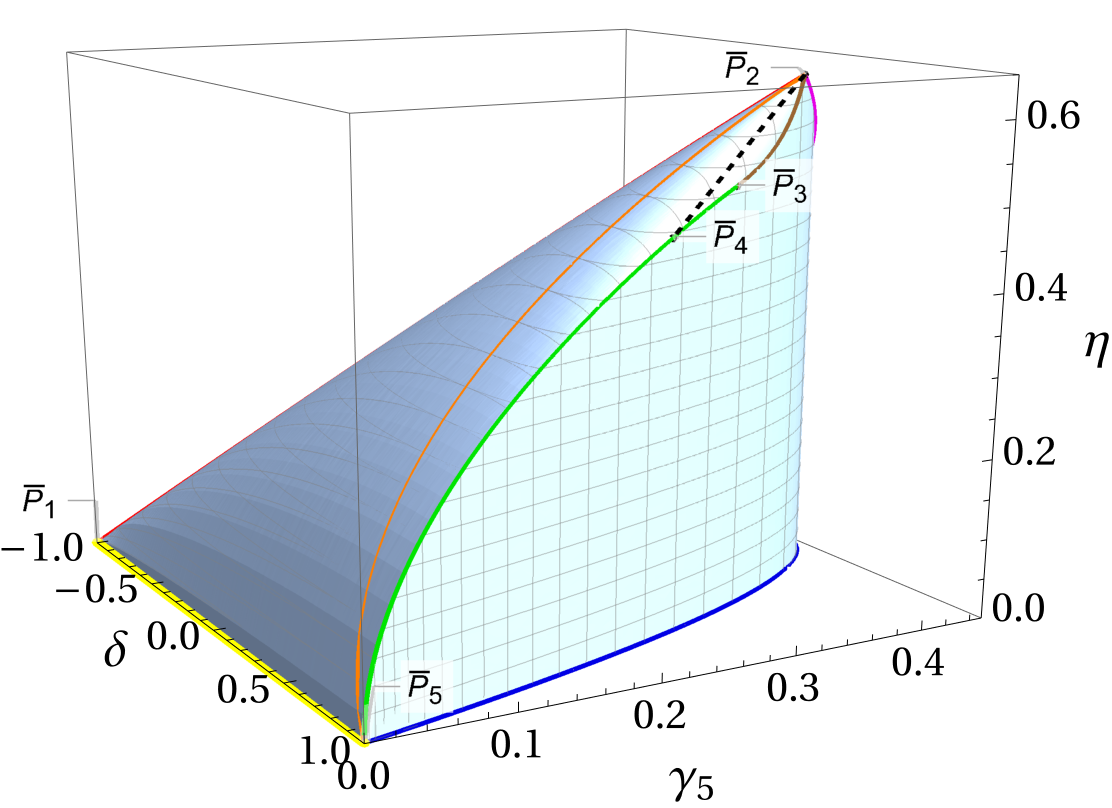}
  \end{center}
  \caption{Two perspectives of the boundary of phase space
    for case $Y=1/2$ with reflection symmetry.
    The points and lines defined in the text are explicitly displayed.}
  \label{fig:Y12-3D}
\end{figure}

\begin{figure}[h!]
  \begin{center}
    \includegraphics[width=1.0\textwidth]{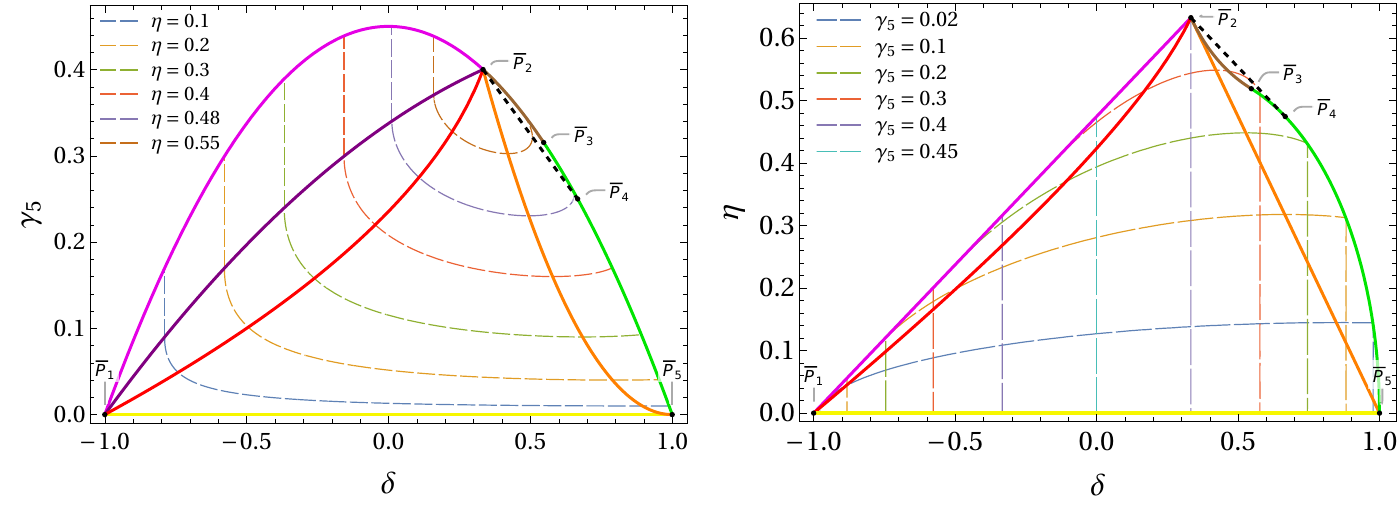}
  \end{center}
  \caption{Left panel:
    the projection of phase space for case $Y=1/2$ with reflection symmetry
    onto the $\gamma_5$ \textit{vs.}\ $\delta$ plane.
    The blue line,
    which is not displayed,
    coincides with the magenta and green--brown lines,
    just as the whole sheet~1 of the boundary of phase space.
    Sheet~2 is between the magenta and purple lines.
    Sheet~3 is among the purple,
    yellow,
    and green--brown lines.
    Sheet~4 is the whole area under the parabola.
    Dashed lines of constant
    $\eta$
    are marked in various colours.
    Right panel:
    the projection of the same phase space
    onto the
    $\eta$
    \textit{vs.}\ $\delta$ plane.
    The blue line,
    which is not displayed,
    coincides with the yellow line,
    just as the whole sheet~4 of the boundary of phase space.
    The purple line,
    which is not displayed,
    coincides with the magenta line,
    just as the whole sheet~2 of the boundary of phase space.
    Sheet~3 is the whole area under the magenta line,
    black dashed line,
    and part of the green segment;
    sheet~1 is the area under the magenta and green--brown lines.
    Dashed lines of constant $\gamma_5$ are marked in various colours.
  }
  \label{fig:Y12-proj}
\end{figure}

\paragraph{Figures:}
The surface that bounds phase space is displayed in Fig.~\ref{fig:Y12-3D}
from two different perspectives.
In Fig.~\ref{fig:Y12-proj} we display the projection of that surface
on the $\gamma_5$ \textit{vs.}\ $\delta$ plane;
in this projection,
the blue line---which is not displayed in Fig.~\ref{fig:Y12-proj}---coincides
with the magenta line for $\delta < 1/3$
and with the green--brown line for $\delta > 1/3$.

\paragraph{Black dashed line:}
Equation $Q_4 = 0$ in general produces six solutions for $\gamma_5$,
but only two of those solutions are relevant for bounding the phase space. 
For $0 \le \eta^2 \le 9/40$,
sheet~3 is formed by solution $\gamma_5^{(1)}\left(\delta, \eta\right)$;
for $9/40 \le \eta^2 \le 2/5$,
sheet~3 is formed
partly by solution $\gamma_5^{(1)}\left(\delta, \eta\right)$
and partly by solution $\gamma_5^{(2)}\left(\delta, \eta\right)$;
those two solutions are separated by,
and coincide,
at the black dashed line.
That line is given by
\bs
\ba
\gamma_5 &=& \frac{11 - 9 \delta}{20},
\\
\eta &=& \frac{5 - 3 \delta}{2 \sqrt{10}},
\ea
\es
for $1/3 \le \delta \le 2/3$.
The black dashed line starts at $\overline P_2$ and ends at $\overline P_4$,
where it meets the green segment of the green--brown line.
As seen in Fig.~\ref{fig:Y12-proj},
the black dashed line is the locus of the points at the boundary of phase space
with maximal $\delta$ for a fixed given
$\eta$,
provided that
$\eta > 3 \left/ \left( 2 \sqrt{10} \right) \right.$.

\paragraph{Boundedness from below:}
The necessary and sufficient conditions
for the potential in Eq.~\eqref{pot2} to be BFB are conditions~\eqref{29a},
\eqref{300} and,
besides~\cite{kannike0},
\be
\lambda_3 + \frac{3}{4}\, \lambda_4 \delta - \left| \chi \right| \eta
+ \sqrt{\lambda_1 \left( \lambda_2 + 2 \lambda_5 \gamma_5 \right)} \ge 0
\label{ciopd}
\ee
must hold for all possible values of $\delta$,
$\gamma_5$,
and $\eta$.
Since the term $\left| \chi \right| \eta$ appears in condition~\eqref{ciopd}
with negative sign,
what matters in practice is the maximum possible $\eta$
for each value of the pair $\left( \delta, \gamma_5 \right)$.
So,
looking at
Fig.~\ref{fig:Y12-3D},
what matters are sheets~2 and~3 of the bounding surface.
On sheet~2,
$\eta$ is a function only of $\delta$ while $\gamma_5$ increases
monotonously when going from the purple line to the magenta line.
One therefore has two further necessary conditions for $V_4$ to be BFB,
namely
\bs
\label{trte}
\ba
\lambda_3 + \frac{3}{4}\, \lambda_4 \delta
- \left| \chi \right| \frac{3 \left( 1 + \delta \right)}{2 \sqrt{10}}
+ \sqrt{\lambda_1 \left[ \lambda_2 + \frac{9}{10}\, \lambda_5
    \left( 1 - \delta^2 \right) \right]} &>& 0,
\label{eq_magenta}
\\
\lambda_3 + \frac{3}{4}\, \lambda_4 \delta
- \left| \chi \right| \frac{3 \left( 1 + \delta \right)}{2 \sqrt{10}}
+ \sqrt{\lambda_1 \left[ \lambda_2 + \frac{9}{40}\, \lambda_5
    \left( 3 - \delta \right) \left( 1 + \delta \right) \right]} &>& 0
\label{eq_purple}
\ea
\es
must hold $\forall \delta \in \left[ -1, 1/3 \right]$.
In practice one finds that condition~\eqref{eq_purple},
relative to the purple line,
is superfluous;
condition~\eqref{eq_magenta},
that follows from the magenta line,
does matter.

\paragraph{Relevance of the orange line and green segment:}

We have randomly generated $10^6$ sets of parameters
$\lambda_1, \ldots, \lambda_5, \chi$
and we have explicitly minimized $V_4$ for each of those sets.
We have found that the BFB conditions obtained in this way
almost coincide with the ones that one finds if,
besides conditions~\eqref{29a},
\eqref{300},
and~\eqref{trte},
one just imposes condition~\eqref{ciopd}
both
on the orange line,
\textit{viz.}
\be
\lambda_3 + \frac{3 \lambda_4 \delta}{4}
- \frac{3 \left| \chi \right| \left( 1 - \delta \right)}{\sqrt{10}}
+ \sqrt{\lambda_1 \left[ \lambda_2
    + \frac{9 \lambda_5 \left( 1 - \delta \right)^2}{10} \right]}
\ge 0 \quad \forall \delta \in \left[ \frac{1}{3}, 1 \right],
\label{orangeL}
\ee
and on the green segment,
\textit{viz.}
\be
\label{greenL}
\lambda_3 + \frac{3 \lambda_4 \delta}{4}
- \frac{3 \left| \chi \right|
  \sqrt{\left( 3 \delta + 1 \right) \left( 1 - \delta \right)}}{2 \sqrt{10}}
+ \sqrt{\lambda_1 \left[ \lambda_2
    + \frac{9 \lambda_5 \left( 1 - \delta^2 \right)}{10} \right]}
\ge 0 \quad \forall \delta \in \left[
  \frac{1 + 2 \sqrt{2}}{7}, 1 \right].
\ee
Only seven quartic potentials were incorrectly identified as BFB in this way.
This occurred because the region
of sheet~3
between the red line and the green--brown line is convex,
and these lines are therefore unable
to correctly locate the minimum for some potentials.
Our numerical analysis showed that
one needs
to re-scan the potentially mis-identified pottentials
by using $10^3$ sets of parameters $\left( \delta, \gamma_5, \eta \right)$
uniformly distributed on sheet~3 between the red and green--brown lines;
this additional procedure allow
one to exactly reproduce the BFB $V_4$'s obtained by the direct minimization.

\paragraph{Procedure:} Our practical procedure
for checking the boundedness-from-below of $V_4$
consists of the following steps.
\begin{enumerate}
\item The necessary conditions~\eqref{29a} and~\eqref{300} are checked.
\item The condition~\eqref{ciopd} is tested
  at the special points~\eqref{points_Y12}.
\item Conditions~\eqref{eq_magenta},
  \eqref{orangeL},
  and~\eqref{greenL} are enforced.
\item The selected points are scanned along the surface $Q_4=0$
  between the red and green--brown lines.
\end{enumerate}

\paragraph{Efficiency of the method:}
In order to demonstrate the efficiency of our method
we have compared it with the brute-force minimization of $V_4$,
following the same procedure as in section~\ref{sec:3/2}.
We randomly generated $10^6$ sets of couplings
$\lambda_1, \ldots \lambda_5, \chi$.
Then:
\begin{itemize}
\item The direct minimization of the $10^6$ $V_4$'s required 62\,157 seconds
  and yielded 381\,910 bounded from below $V_4$'s.
\item Our proposed method\footnote{Since the lines
in Eqs.~\eqref{trte}–\eqref{greenL}
depend on the single parameter $\delta$ and are continuously differentiable,
we have employed for them an efficient bisection minimization method
instead of a direct scan.}
completed the same task in 36.6 seconds,
identifying exactly the same 381\,910 BFB potentials.
\end{itemize}
This comparison demonstrates that our method is approximately 1\,700 times
faster than the brute-force minimization of $V_4$.

\section{Conclusions and discussion}
\label{sec:conclusions}

The important
findings in this work were the following:
\begin{itemize}
\item For the two models studied,
  there are \emph{analytical} equations for the boundaries
  of the phase spaces.
  Besides the trivial equations
  $\epsilon = 0$ in Section~\ref{sec:3/2}
  and $\eta = 0$ in Section~\ref{sec:1/2},
  there are the equations
  \be
  Q_1 = 9 \left( 1 - \delta^2 \right) - 20 \gamma_5 = 0
  \ee
  in both Sections~\ref{sec:3/2} and~\ref{sec:1/2},
  and
  \be
  Q_3 = 9 \left( 1 + \delta \right)^2 - 40 \eta^2 = 0
  \ee
  in Section~\ref{sec:1/2}.
  \emph{Most important of all,
  there are equations $Q_2 = 0$ in Section~\ref{sec:3/2}
  and $Q_4 = 0$ in Section~\ref{sec:1/2}},
  where $Q_2$ and $Q_4$ are polynomial functions
  explicitly displayed in Appendix~\ref{App:expressions}.
  The latter equations have several solutions;
  usually,
  either two or three of
  those
  solutions are relevant for
  the boundaries
  of the phase spaces,
  while the other solutions are to be discarded.
\item It is necessary to assess the convexity or concavity
  of the (upper) boundary of phase space\footnote{The importance
    of the convexity or concavity of phase space
    for the determination of the minima of the scalar potential
    was emphasized in Ref.~\cite{ivanov}.
    Convexity or concavity of phase space is also important
    for constructing sufficient and/or necessary conditions
    for boundedness-from-below of the potential;
    see for instance Ref.~\cite{other2},
    where the importante of constructing a `convex hull' of phase space
    was emphasized.}
  in order to understand whether
  one must scan that surface fully,
  or just a few lines thereon,
  when studying the boundedness-from-below of the scalar potential.
  Namely,
  convex boundaries require a full scan
  (and the more convex they are,
  the more detailed that scan must be),
  while for concave or indefinite-concavity boundaries
  the scan over a few lines suffices.
  In order to correctly assess the convexity or concavity
  one must write the quartic potential
  as a \emph{linear} combination of the phase-space parameters,
  \textit{viz.}\ of $\delta$,
  $\gamma_5$,
  and either $\epsilon$ or $\eta$,
  \textit{cf.}\ Eqs.~\eqref{pot1} and~\eqref{pot2}---and employ just those
  parameters---not, say, their squares---in the procedure
  explained in Appendix~\ref{App:concavity}.
\item The bounded-from-below conditions on the quartic parts
  of the scalar potentials
  mostly
  require,
  in practice,
  numerical scans over a few
  \emph{lines}---not surfaces\footnote{In the case in Section~\ref{sec:1/2}
  a scan over the surface with $Q_4 = 0$ is finally needed
  in order to eliminate a few unbounded-from-below potentials.}---at
  the boundary of the phase space
  of each model.\footnote{We formulate the
  \emph{hypothesis}
  that this
  happens
  because of the concave,
  or partly concave,
  nature of the relevant stretches of the boundary of phase space.
  When that boundary is distinctly convex---an example is displayed
  in Appendix~\ref{App:more}---numerical scans just over a few lines
  are
  probably not
  enough.}
  We have explicitly given those lines for each of the
  two
  models.
\end{itemize}

We have found,
by giving random values to all the scalar fields,
that the phase spaces that one obtains are the same,
irrespective of whether one allows the fields
to be complex---as they in general are---or one restricts all of them
to be real.
The reason why this is so remains to be investigated.

If the fields are taken to be real,
then in both models that we have studied there are three phase-space coordinates
($\delta$,
$\gamma_5$,
and either $\epsilon$ or $\eta$)
to be determined by three relative field values,
\textit{viz.}\ $c/f$,
$d/f$,
and $e/f$ in the gauge where $a=0$.
This makes the models relatively simple,
because one just needs to set one determinant equal to zero;
no \textit{ad hoc}\/ choices of field directions are necessary.
In other models that one may want to study
things most likely will be more
complicated---with a number of relative field values larger than the one
of phase-space coordinates.
This is another issue that remains to be investigated.

It must be stressed that the analytical methods that we have utilized
did not dispense with the confirmation of all the results
by giving random values to all the scalar fields
and observing the phase spaces that get generated in this way.
All our analytical phase spaces have been numerically confirmed
in this way.

Usually,
checking the BFB conditions
through a brute-force,
high-precision minimization of
the quartic part of the potential
demands substantial computational resources. 
Scanning a few conditions along the specific lines
identified in our analyses
accelerates
the computation by several hundred times. 
Employing efficient minimization techniques
along those lines
may improve the computational performance
by up to three orders of magnitude;
that was a significant achievement of this work.

The expressions in Appendix~\ref{App:expressions}, 
together with computational examples in {\tt Mathematica} notebook files, 
are available at \href{https://github.com/jurciukonis/SM-quadruplet}
{\tt https://github.com/jurciukonis/SM-quadruplet}, 
enabling the reproduction of the results presented in this work.

\vspace*{3mm}

\paragraph{Acknowledgements:}
The work of D.J.\ received funding from the Research Council of Lithuania
(LMT) under Contract No.\ S-CERN-24-2.
Part of the computations were performed using the infrastructure
of the Lithuanian Particle Physics Consortium
in the framework of the agreement No.\ VS-13 of Vilnius University with LMT.
The work of L.L.\ is supported
by the Portuguese Foundation for Science and Technology (FCT)
through projects UIDB/00777/2020 and UIDP/00777/2020,
and by the PRR (Recovery and Resilience Plan),
within the scope of the investment
``RE-C06-i06 - Science Plus Capacity Building",
measure ``RE-C06-i06.m02 - Reinforcement of Financing
for International Partnerships in Science, Technology and Innovation
of the PRR,"
under the project with reference 2024.01362.CERN.
L.L.\ was furthermore supported by projects CERN/FIS-PAR/0002/2021
and CERN/FIS-PAR/0019/2021.

\newpage

\begin{appendix}

\setcounter{equation}{0}
\renewcommand{\theequation}{A\arabic{equation}}

\section{The expressions of $Q_2$ and $Q_4$}
\label{App:expressions}

\paragraph{$Q_2$:}
For the quantity $Q_2$ in Eq.~\eqref{q2} one has
\be
Q_2 = \sum_{n=0}^6 b_n \left( 10 \gamma_5 \right)^n.
\label{eq_C=0}
\ee
The coefficients $b_0, \ldots, b_6$
are functions of $\delta$ and $\epsilon^2$:
\bs
\ba
b_0 &=& 3^8 \left[ \left( 1 + \delta \right)^3 - 8 \epsilon^2 \right]^2
\left[ \left( 127 + 225 \delta + 405 \delta^2 + 243 \delta^3\right)^2
  \right. \\ & &
  - 4 \left( 28\,795 + 164\,619 \delta + 281\,070 \delta^2
  + 253\,206 \delta^3 + 185\,895 \delta^4 + 98\,415 \delta^5 \right)
  \epsilon^2
  \hspace*{7mm}
  \\ & &
  + 12 \left( 49\,331 + 167\,724 \delta + 174\,546 \delta^2
  + 65\,772 \delta^3 + 54\,675 \delta^4 \right) \epsilon ^4
  \\ & & \left.
  - 4\,096 \left( 292 + 585 \delta + 54 \delta^2
  + 81 \delta^3 \right) \epsilon ^6
  + 1\,048\,576 \epsilon ^8
  \vphantom{\left( 127 + 225 \delta + 405 \delta^2 + 243 \delta^3\right)^2}
  \right],
\ea
\es
\bs
\ba
b_1 &=& - 2 \times 3^6 \left[ \left(1 + \delta\right)^3
  \left( 127 + 225 \delta + 405 \delta^2 + 243 \delta^3 \right)
  \left( \vphantom{729 \delta^4} 1\,129
  \right. \right. \\ & & \left.
  + 2\,868 \delta + 4\,614 \delta^2 + 3\,348 \delta^3 + 729 \delta^4 \right)
  \\ & &
  - 8 \left( 241\,961 + 1\,921\,725 \delta + 6\,748\,761 \delta^2
  + 14\,553\,795 \delta^3 + 20\,815\,167 \delta^4
  \right. \\ & & \left.
  + 20\,846\,649 \delta^5 + 14\,579\,811 \delta^6 + 6\,344\,001 \delta^7
  + 1\,303\,452 \delta^8 + 39\,366 \delta^9 \right) \epsilon^2
  \\ & &
  + 4 \left( 3\,673\,191 + 24\,261\,588 \delta + 77\,591\,040 \delta^2
  + 128\,209\,796 \delta^3 + 127\,686\,438 \delta^4
  \hspace*{7mm}
  \right. \\ & & \left.
  + 86\,310\,396 \delta^5 + 40\,192\,344 \delta^6 + 8\,427\,564 \delta^7
  - 639\,333 \delta^8 \right) \epsilon^4
  \\ & &
  - 16 \left( 3\,786\,799 + 23\,774\,769 \delta + 51\,666\,021 \delta^2
  + 56\,099\,983 \delta^3 + 33\,353\,169 \delta^4
  \right. \\ & & \left.
  + 14\,415\,495 \delta^5 + 2\,899\,899 \delta^6 - 372\,519 \delta^7
  \right) \epsilon^6
  \\ & &
  + 512 \left( 350\,855 + 1\,404\,999 \delta + 1\,906\,131 \delta^2
  + 844\,173 \delta^3 + 304\,122 \delta^4
  \right. \\ & & \left.
  + 43\,680 \delta^5 - 6\,624 \delta^6 \right) \epsilon^8
  \\ & & \left.
  - 262\,144 \left( 1\,041 + 2\,439 \delta + 546 \delta^2
  + 164 \delta^3 \right) \epsilon^{10}
  + 201\,326\,592 \epsilon^{12} \right], 
\ea
\es
\bs
\ba
b_2 &=& 3^4 \left[ 1\,998\,033 + 10\,659\,240 \delta
  + 31\,114\,620 \delta^2 + 58\,455\,512 \delta^3 + 75\,085\,254 \delta^4
  \right. \\ & &
  + 65\,745\,432 \delta^5 + 37\,096\,380 \delta^6 + 12\,159\,720 \delta^7
  + 1\,791\,153 \delta^8
  \\ & &
  - 12 \left( 1\,405\,269 + 10\,375\,407 \delta + 27\,924\,145 \delta^2
  + 45\,055\,699 \delta^3 + 45\,386\,607 \delta^4
  \right. \\ & & \left.
  + 28\,978\,749 \delta^5 + 10\,862\,667 \delta^6
  + 1\,709\,505 \delta^7 \right) \epsilon^2
  \\ & &
  + 4 \left( 26\,754\,009 + 143\,143\,614 \delta + 389\,331\,183 \delta^2
  + 492\,368\,452 \delta^3 + 327\,588\,327 \delta^4
  \hspace*{10mm}
  \right. \\ & & \left.
  + 112\,833\,054 \delta^5 + 23\,517\,297 \delta^6 \right) \epsilon^4
  \\ & &
  - 64 \left( 4\,988\,475 + 29\,262\,861 \delta + 54\,167\,994 \delta^2
  + 46\,168\,094 \delta^3 
  \right. \\ & & \left.
  + 12\,881\,115 \delta^4 + 2\,068\,389 \delta^5 \right) \epsilon^6
  \\ & &
  + 1\,536 \left( 524\,649 + 1\,964\,682 \delta + 2\,513\,849 \delta^2
  + 517\,512 \delta^3 + 33\,072 \delta^4 \right) \epsilon^8
  \\ & & \left.
  - 262\,144 \left( 4\,056 + 10\,269 \delta + 1\,200 \delta^2
  - 32 \delta^3 \right) \epsilon^{10}
  + 805\,306\,368 \epsilon^{12} \right],
\ea
\es
\bs
\ba
b_3 &=& - 2^5 \times 3^2 \left[ 219\,250 + 6 \delta \left( 133\,498
  + 288\,469 \delta + 369\,420 \delta^2
  \right. \right. \\ & & \left.
  + 298\,701 \delta^3 + 139\,482 \delta^4 + 25\,515 \delta^5 \right)
  \\ & &
  - 2 \left( 573\,229 + 3\,817\,557 \delta + 7\,002\,354 \delta^2
  + 7\,473\,690 \delta^3 
  \right. \\ & & \left.
  + 4\,385\,745 \delta^4 + 1\,289\,601 \delta^5 \right) \epsilon^2
  \\ & &
  + \left( 6\,285\,479 + 23\,724\,876 \delta + 49\,111\,578 \delta^2
  + 39\,010\,284 \delta^3 + 14\,624\,631 \delta^4 \right) \epsilon^4
  \hspace*{7mm}
  \\ & &
  - 2 \left( 6\,247\,069 + 29\,947\,395 \delta + 37\,835\,679 \delta^2
  + 24\,267\,897 \delta^3 \right) \epsilon^6
  \\ & &
  + 32 \left( 752\,119 + 2\,076\,849 \delta + 2\,461\,170 \delta^2 \right)
  \epsilon^8 - 16\,384 \left( 1\,316 + 3\,621 \delta \right) \epsilon^{10}
  \\ & & \left.
  + 16\,777\,216 \epsilon^{12} \right],
\ea
\es
\bs
\ba
b_4 &=& 2^8 \left[ 52\,006 + 18 \delta \left( 6\,428
  + 9\,618 \delta + 6\,588 \delta^2 + 2\,187 \delta^3 \right)
  \right. \\ & &
  - 4 \left( 37\,645 + 222\,291 \delta + 212\,463 \delta^2
  + 105\,705 \delta^3 \right) \epsilon^2
  \\ & &
  + 48 \left( 14\,963 + 30\,762 \delta + 38\,871 \delta^2 \right) \epsilon^4
  \\ & & \left.
  - 28 \left( 28\,411 + 88\,407 \delta \right) \epsilon^6
  + 1\,024\,009 \epsilon^8 \right],
\ea
\es
\ba
b_5 &=& - 2^{13} \left[ 2 \left( 89 + 90 \delta + 81 \delta^2 \right)
  - 2 \left(103 + 531 \delta \right) \epsilon^2 + 811 \epsilon^4 \right],
\ea
\ba
b_6 &=& 2^{16}.
\ea

\paragraph{$Q_4$:}
For the quantity $Q_4$ in Eq.~\eqref{jvfdodo} one has
\be
Q_4 = \sum_{n=0}^6 d_n \gamma_5^n.
\label{eq_D=0}
\ee
The coefficients $d_0, \ldots, d_6$
are functions of $\delta$ and $\eta^2$:
\ba
d_0 &=& \eta^4 \left[ 104\,976 R^4
+ 3\,732\,480 \left( 7 - 10 \delta - \delta^2 \right) \eta^2
+ 33\,177\,600\, \eta^4 \right],
\ea
\bs
\ba
d_1 &=& - \eta^2 \left[
  52\,488 R^5 S
  + 466\,560 R
  \left( 84 - 61 \delta - 38 \delta^2 - 5 \delta^3 \right) \eta^2
  \right. \\ & & \left.
  + 8\,294\,400 \left( 29 - 19 \delta - 4 \delta^2 \right) \eta^4
  + 147\,456\,000\, \eta^6 \right],
\ea
\es
\bs
\ba
d_2 &=& 6\,561 R^6 S^2
+ 29\,160 R^2
\left( 513 + 48 \delta - 400 \delta^2 - 76 \delta^3 - 13 \delta^4 \right)
\eta^2
\\ & &
+ 259\,200 \left( 2\,241 - 1\,474 \delta - 785 \delta^2 + 160 \delta^3
+ 30 \delta^4 \right) \eta^4
\\ & &
+ 4\,608\,000 \left( 149 - 36 \delta - 14 \delta^2 \right) \eta^6
+ 163\,840\,000\, \eta^8,
\ea
\es
\bs
\ba
d_3 &=& - 14\,580 R^3 S \left( 117 - 6 \delta - 59 \delta^2 \right)
\\ & &
- 64\,800 \left( 8\,253 - 3\,378 \delta - 5\,260 \delta^2
+ 1\,202 \delta^3 + 463 \delta^4 \right) \eta^2
\\ & &
- 576\,000 \left( 2116 - 623 \delta - 487 \delta^2 \right) \eta^4
- 624\,640\,000\, \eta^6,
\ea
\es
\bs
\ba
d_4 &=& 900 \left[ 9 \left( 20\,601 - 1\,404 \delta - 18\,378 \delta^2
+ 2\,756 \delta^3 + 3\,225 \delta^4 \right)
\right. \\ & & \left.
+ 80 \left( 13\,119 - 2\,770 \delta - 4\,945 \delta^2 \right) \eta^2
+ 1\,025\,600\, \eta^4 \right],
\ea
\es
\ba
d_5 &=& - 2^{11} \times 5^3
\left[ 9 \left( 117 - 6 \delta - 59 \delta^2 \right)
+ 2\,440\, \eta^2 \right],
\ea
\ba
d_6 &=& 2^{18} \times 5^4.
\ea
Here,
\bs
\ba
R &=& 3 - \delta,
\\
S &=& 1 + \delta.
\ea
\es

\newpage

\setcounter{equation}{0}
\renewcommand{\theequation}{B\arabic{equation}}

\section{Concavity or convexity of the phase-space boundary}
\label{App:concavity}

If a surface is the topological boundary
of a set of points in three dimensions,
then that surface is said to be concave (or `concave up') at a point
if it bends toward the outer-pointing surface normal at that point;
the surface is convex (or `concave down') if it bends away
from the outer-pointing normal.

Let $z = f(x, y)$ represent the surface bounding \emph{above}
a volume in $\left( x, y, z \right)$ space.
Then the local concavity
is characterized by the definiteness of the Hessian matrix
\be
H =
\left( \begin{array}{cc}
  \displaystyle{\frac{\partial^2 f}{\partial x^2}} &
  \displaystyle{\frac{\partial^2 f}{\partial x \partial y}}
  \\*[3mm]
  \displaystyle{\frac{\partial^2 f}{\partial x \partial y}} &
  \displaystyle{\frac{\partial^2 f}{\partial y^2}}
\end{array} \right)
: =
\left( \begin{array}{cc}
  f_{xx} &  f_{xy} \\
  f_{xy} & f_{yy} 
\end{array} \right).
\label{Hess}
\ee
\begin{itemize}
\item The surface is locally concave if the Hessian is positive semidefinite,
  \textit{viz.}\ if $\det H \ge 0$,
  $f_{xx} \ge 0$,
  and $f_{yy} \ge 0$.
\item The surface is locally convex if the Hessian is negative semidefinite,
  \textit{viz.}\ if $\det H \ge 0$,
  $f_{xx} \le 0$,
  and $f_{yy} \le 0$.
\item The surface is neither concave nor convex
  if the Hessian is indefinite,\footnote{In this case,
  the surface may be concave in one direction
  but convex in another direction at the same point.}
  \textit{viz.}\ if $\det H \le 0$.
\end{itemize}
The boundary between regions of definite concavity
(either concave up or concave down)
and indefinite concavity may be found by solving the equation $\det H = 0$.

In our specific cases,
the concave and convex regions of the surface bounding the phase space
from above
may be identified by scanning the surface over the parameters
$\delta$ and $\gamma_5$ and using the following procedure:
\begin{enumerate}
\item Obtain analytical solutions of $Q_i = 0$ for
  $i = 2, 4, 5$,
  with respect to the relevant parameters---either $\epsilon$,
  $\eta$, or $\varepsilon$,
  respectively.
\item Determine which of the solutions
  corresponds to the surface at the given point.
\item Compute the Hessian matrix at that point
  by double-differentiating the relevant solution
  with respect to $\delta$ and $\gamma_5$.
\item Evaluate the concavity conditions at the given point,
  \textit{viz.}\ determine the signs of $\det H$,
  $f_{xx}$,
  and $f_{yy}$.
\end{enumerate}

\newpage

\setcounter{equation}{0}
\renewcommand{\theequation}{C\arabic{equation}}

\section{One further case}
\label{App:more}

\paragraph{Potential:}
If the doublet $\Phi$ and the quadruplet $\Xi$ have the same hypercharge
and if there is no symmetry under $\Xi \to - \Xi$,
then in general
\be
V_{4,\mathrm{extra}} = \frac{\chi}{2}\, T_3^\dagger T_1
+ \frac{\nu}{2}\, T_4^\dagger T_3
+ \frac{\iota}{2}\, T_4^\dagger T_1
+ \mathrm{H.c.},
\label{9v0f00f}
\ee
where
$T_1$ is the $SU(2)$ triplet in Eq.~\eqref{tir1},
$T_3$ is the $SU(2)$ triplet in Eq.~\eqref{tir3},
and
\be
T_4 = \left( \Xi \otimes \Phi \right)_\mathbf{3}
= \frac{1}{2} \left( \begin{array}{c}
  \sqrt{3}\, b c - a d \\
  \sqrt{2} \left( b d - a e \right) \\
  b e - \sqrt{3}\, a f
\end{array} \right)
\label{tir4}
\ee
is one further $SU(2)$ triplet.
In this Appendix we follow Ref.~\cite{kannike} and assume $\chi = \iota = 0$,
even if this is inconsistent because there is no symmetry that enforcs this
and therefore nonzero $\chi$ and $\iota$ are needed for renormalization.

\paragraph{Parameter $\varepsilon$:}
The dimensionless parameter
\be
\varepsilon \equiv \frac{\left| T_4^\dagger T_3 \right|}{\sqrt{F_1^3 F_2}}
\ee
is non-negative by definition.
Then (with $\chi = \iota = 0$),
\be
V_{4,\mathrm{extra}} = \left| \nu \right| \varepsilon\, \sqrt{F_1^3 F_2}\,
\cos \left( \arg \nu + \arg T_4^\dagger T_3 \right),
\ee
so that
\be
\frac{V_4}{F_2^2} =
\frac{\lambda_1}{2} \left( \sqrt{r} \right)^4
+ \left| \nu \right| \varepsilon
\cos \left( \arg \nu + \arg T_4^\dagger T_3 \right) \left( \sqrt{r} \right)^3
+ \left( \lambda_3 + \frac{3}{4}\, \lambda_4 \delta \right)
\left( \sqrt{r} \right)^2
+ \frac{\lambda_2 + 2 \lambda_5 \gamma_5}{2}.
\label{c5}
\ee

\paragraph{Gauge $a=0$:}
In the gauge where $a=0$ throughout space--time,
$T_4^\dagger T_3 = \left( F_1 / 2 \right) b e^\ast$ and therefore
\be
\varepsilon^2 = \frac{E}{4 \left( F + E + D + C \right)}.
\label{vmif00}
\ee
From Eqs.~\eqref{deltaa=0} and~\eqref{vmif00} we note that
\bs
\ba
3 \left( 1 + \delta \right) - 16 \varepsilon^2 &=&
\frac{6 F + 2 D}{F + E + D + C},
\\
3 \left( 1 - \delta \right) - 8 \varepsilon^2 &=&
\frac{6 C + 4 D}{F + E + D + C}
\ea
\es
are both non-negative.
There are thus two upper bounds on
$\varepsilon^2$:
\bs
\ba
\varepsilon^2 &\le& \frac{3 \left( 1 + \delta \right)}{16},
\label{b1}
\\
\varepsilon^2 &\le& \frac{3 \left( 1 - \delta \right)}{8}.
\label{b2}
\ea
\es
It follows from them
that the maximum possible value of $\varepsilon^2$ is $1/4$
and that it is attained when $\delta = 1/3$,
\textit{viz.}\ at the point that we shall call $\widehat P_3$
in Eq.~\eqref{widehatP3}.

\paragraph{Boundary:}
Using Eqs.~\eqref{15} and
\be
\varepsilon^2 = \frac{z^2}{4},
\ee
we write the condition
\be
\det \left( \begin{array}{ccc}
  \displaystyle{\frac{\partial \delta}{\partial x}} &
  \displaystyle{\frac{\partial \delta}{\partial y}} &
  \displaystyle{\frac{\partial \delta}{\partial z}} \\*[3mm]
  \displaystyle{\frac{\partial \gamma_5}{\partial x}} &
  \displaystyle{\frac{\partial \gamma_5}{\partial y}} &
  \displaystyle{\frac{\partial \gamma_5}{\partial z}} \\*[3mm]
  \displaystyle{\frac{\partial \varepsilon^2}{\partial x}} &
  \displaystyle{\frac{\partial \varepsilon^2}{\partial y}} &
  \displaystyle{\frac{\partial \varepsilon^2}{\partial z}}
\end{array} \right) = 0.
\ee
We therefrom obtain
\be
\varepsilon^2\, Q_1
Q_5 = 0, \label{Q1Q2}
\ee
where $Q_1$ is the quantity defined in Eq.~\eqref{14a} and
\be
Q_5 = \sum_{n=0}^6 e_n \left(10 \gamma_5 \right)^n.
\label{eq_Q6}
\ee
The coefficients $e_0, \ldots, e_6$
are functions of $\delta$ and $\varepsilon^2$.
Defining
\bs
\ba
\tilde R &=& \delta - 1,
\\
\tilde S &=& 9 \delta + 7,
\ea
\es
one has
\bs
\ba
e_0 &=& - 3^6
\left[ 3 \tilde R \left( \delta + 1 \right)^2 + 32 \varepsilon^2 \right]^2
\left[ 81 \tilde R^2 \tilde S^4
  + 432 \tilde R \tilde S^2 \left( 7 \delta + 9 \right)
  (21 - 5 \delta) \varepsilon^2
  \right. \\ && \left.
  + 576 \left( 56\,841 + 81\,820 \delta + 10\,998 \delta^2
  - 52\,452 \delta^3 - 31\,671 \delta^4 \right) \varepsilon^4
  \right. \\ && \left.
  - 2^{18} \times 3 \left(172 + 183 \delta - 18 \delta^2
  - 81 \delta^3 \right) \varepsilon^6
  + 2^{28} \varepsilon^8
  \vphantom{\tilde R^2 \tilde S^4} \right],
\ea
\es
\bs
\ba
e_1 &=& 2 \times 3^5
\left[ 243 \tilde R^3 \tilde S^3 \left(1 + \delta\right)^2
  \left(39 + 66 \delta + 23 \delta^2 \right)
  \right. \\ &&
  + 2\,592 \tilde R^2 \tilde S
  \left(6\,209 + 26\,546 \delta + 40\,720 \delta^2 + 24\,272 \delta^3
  \right. \\ &&  \left.
  + 1\,773 \delta^4 - 1\,674 \delta^5 + 458 \delta^6 \right) \varepsilon^2
  \\ &&
  + 1\,728 \tilde R
  \left(975\,729 + 3\,617\,501 \delta + 4\,193\,369 \delta^2
  + 606\,141 \delta^3
  \right. \\ && \left.
  - 1\,803\,301 \delta^4 - 995\,505 \delta^5 - 223\,269 \delta^6
  - 79\,209 \delta^7 \right) \varepsilon^4
  \\ &&
  + 9\,216 \left( 1\,506\,843 + 2\,982\,015 \delta
  + 68\,969 \delta^2 - 2\,724\,639 \delta^3
  \right. \\ && \left.
  - 1\,269\,563 \delta^4 + 330\,409 \delta^5
  + 203\,799 \delta^6 - 49\,257 \delta^7 \right) \varepsilon^6
  \\ &&
  - 2^{17} \times 3 \left(171\,715 + 238\,925 \delta - 8\,745 \delta^2
  \right. \\ &&  \left.
  - 146\,049 \delta^3 - 38\,502 \delta^4 + 14\,560 \delta^5
  -  6\,624 \delta^6 \right) \varepsilon^8
  \\ && \left.
  + 2^{28} \left( 717 + 637 \delta - 182 \delta^2
  - 164 \delta^3 \right) \varepsilon^{10}
  - 2^{38} \varepsilon^{12}
  \vphantom{\tilde R^2 \tilde S^4} \right],
\ea
\es
\bs
\ba
e_2 &=& 3^3 \left[ 243 \tilde R^2 \tilde S^2
  \left(2\,481 + 8\,220 \delta + 9\,350 \delta^2
  + 4\,060 \delta^3 + 465 \delta^4 \right)
  \right. \\ &&
  +  1\,296 \tilde R \left( 640\,633 + 2\,533\,410 \delta
  + 3\,215\,487 \delta^2 + 812\,348 \delta^3
  \right. \\ && \left.
  - 927\,001 \delta^4 - 202\,494 \delta^5
  + 219\,073 \delta^6 \right) \varepsilon^2
  \\ &&
  + 1\,728 \left( 6\,014\,521 + 13\,113\,034 \delta
  + 940\,951 \delta^2 - 11\,539\,124 \delta^3
  \right. \\ && \left.
  - 3\,255\,977 \delta^4 + 1\,511\,370 \delta^5
  - 493\,319 \delta^6 \right) \varepsilon^4
  \\ &&
  - 2^{12} \times 3^2 \left( 1\,990\,309 + 3\,074\,189 \delta
  - 369\,486 \delta^2
  \right. \\ && \left.
  - 1\,636\,950 \delta^3
  - 91\,351 \delta^4 + 211\,785 \delta^5 \right) \varepsilon^6
  \\ &&
  + 2^{17} \times 3 \left( 790\,995 + 797\,218 \delta - 103\,565 \delta^2
  - 178\,168 \delta^3 + 73\,616 \delta^4 \right) \varepsilon^8
  \\ && \left.
  - 2^{28} \left( 2\,904 + 2\,051 \delta - 400 \delta^2
  + 32 \delta^3 \right) \varepsilon^{10}
  + 2^{40} \varepsilon^{12}
  \vphantom{\tilde R \tilde S^2} \right],
\ea
\es
\bs
\ba
e_3 &=& 2^6 \left[ 729 \tilde R \tilde S
  \left(641 + 1\,820 \delta + 1\,510 \delta^2
  + 220 \delta^3 - 95 \delta^4 \right)
  \right. \\ &&
  + 972 \left( 74\,101 + 171\,935 \delta
  + 26\,162 \delta^2
  \right. \\ && \left.
  - 148\,706 \delta^3
  - 36\,503 \delta^4 + 44\,083 \delta^5 \right) \varepsilon^2
  \\ &&
  - 648 \left( 1\,097\,753 + 1\,761\,596 \delta - 156\,618 \delta^2
  - 756\,868 \delta^3 + 118\,521 \delta^4 \right) \varepsilon^4
  \\ &&
  + 1\,728 \left( 2\,284\,741 + 2\,537\,353 \delta - 580\,705 \delta^2
  - 622\,573 \delta^3 \right) \varepsilon^6
  \\ &&
  - 2^{12} \times 3^2
  \left( 362\,107 + 203\,059 \delta - 58\,878 \delta^2 \right)
  \varepsilon^8
  \\ &&  \left.
  + 2^{23} \times 3 \left( 980 + 491 \delta \right) \varepsilon^{10}
  - 2^{35} \varepsilon^{12}
  \vphantom{\tilde R \tilde S} \right],
\ea
\es
\bs
\ba
e_4 &=& 2^9 \left[ 81 \left( 723 + 1620 \delta + 530 \delta^2
  - 620 \delta^3 - 205 \delta^4 \right)
  \right. \\ &&
  - 216 \left( 4\,331 + 7\,303 \delta - 271 \delta^2
  - 3\,683 \delta^3 \right) \varepsilon^2
  \\ &&
  + 1\,152 \left( 5\,757 + 5\,450 \delta - 1\,295 \delta^2 \right)
  \varepsilon^4
  \\ && \left.
  - 1\,152 \left( 19\,977 + 14\,623 \delta \right) \varepsilon^6
  + 49\,280\,128 \varepsilon^8
  \vphantom{\delta^4} \right],
\ea
\es
\ba
e_5 &=& - 2^{14} \left[ 135 + 162 \delta - 9 \delta^2
  - \left( 1\,236 + 1\,356 \delta \right) \varepsilon^2
  + 5\,224 \varepsilon^4
  \right],
\ea
\ba
e_6 &=& 2^{16}.
\ea
It follows from Eq.~\eqref{Q1Q2} that the border of the
$\left( \delta, \gamma_5, \varepsilon \right)$
phase space
has three sheets:
\begin{enumerate}
\item Sheet~1 has equation $Q_1 = 0$.
\item Sheet~2 has equation $\varepsilon = 0$.
\item Sheet~3 has equation
  $Q_5 = 0$.
\end{enumerate}
It should be emphasized that the equation
$Q_5 = 0$
usually yields up to six real solutions for $\varepsilon^2$,
for any given values of $\delta$ and $\gamma_5$.
(That equation may also have non-real solutions.)
However,
not all of those solutions form the border of phase space.
That border must be confirmed by giving random (in general, complex) values
to the fields $c$,
$d$,
$e$,
and $f$ and explicitly constructing the phase space
via
Eqs.~\eqref{g5},
\eqref{deltaa=0},
and~\eqref{vmif00}.

\paragraph{Points:} We define the points
\bs
\label{points_Y12K}
\ba
\widehat P_1: & & \delta = -1,\ \gamma_5 = \varepsilon = 0;
\\
\widehat P_2: & & \delta = 1,\ \gamma_5 = \varepsilon = 0;
\\
\widehat P_3: & & \delta = \frac{1}{3},\ \gamma_5 = \frac{2}{5},\
\varepsilon = \frac{1}{2};
\label{widehatP3}
\\
\widehat P_4: & & \delta = \frac{1}{3},\ \gamma_5 = 0,\
\varepsilon = \frac{1}{3};
\\
\widehat P_5: & & \delta = \frac{1}{9},\ \gamma_5 = \frac{2}{5},\
\varepsilon = 0.
\ea
\es
Points $\widehat P_1$ and $\widehat P_2$ have $Q_1 = \varepsilon =
Q_5 = 0$.
Point $\widehat P_3$ has $Q_1 =
Q_5 = 0$
but $\varepsilon \neq 0$.
Point $\widehat P_5$ has $\varepsilon =
Q_5 = 0$ but $Q_1 \neq 0$.
Point $\widehat P_4$ is solely on sheet~3.

\paragraph{Main lines:}
We define three lines connecting point $\widehat P_1$ to point $\widehat P_2$:
\begin{itemize}
\item The blue line has $Q_1 = \varepsilon = 0$.
\item The purple line has $\varepsilon =
  Q_5 = 0$;
  this gives
  \be
  \gamma_5 = \left\{ \begin{array}{lcl}
    \displaystyle{\frac{9 \left( 1 + \delta \right)^2}{10}} & \Leftarrow &
    \displaystyle{-1 \le \delta \le - \frac{3}{5}},
    \\*[3mm]
    \displaystyle{- \frac{9 \tilde R \tilde S}{160}} & \Leftarrow &
    \displaystyle{- \frac{3}{5} \le \delta \le 1}.
  \end{array} \right.
  \ee
  Point $\widehat P_5$ is on the purple line.
\item The magenta--brown line has $Q_1 =
  Q_5 = 0$.
  It has two segments,
  the magenta one from $\widehat P_1$ to $\widehat P_3$
  and the brown one from $\widehat P_3$ to $\widehat P_2$.
  Both segments have
  $\gamma_5 = \left. 9 \left( 1 - \delta^2 \right) \right/ 20$;
  the magenta segment has
  \be
  \varepsilon^2 = \frac{3 \left( 1 + \delta \right)}{16},
  \label{uw1}
  \ee
  while the brown segment has
  \be
  \varepsilon^2 = \frac{3 \left[ 27
    + 36 \delta - 142 \delta^2 + 68 \delta^3 + 11 \delta^4
    + \sqrt{- \tilde R^5 \left( 23 \delta + 9 \right)^3}
    \right]}{16\,384\, \delta^3}.
  \label{uw2}
  \ee
  At their meeting point $\widehat P_3$ the magenta and brown segments
  have derivatives $\left. \mathrm{d} \varepsilon^2 \right/ \mathrm{d} \delta$
  of opposite sign.
  Notice that the magenta segment materializes the upper bound~\eqref{b1}
  on $\varepsilon^2$.
\end{itemize}

\paragraph{Bounding surface:} The surface bounding phase space
has the topology of the surface of a sphere.
On that sphere lie the two points $\widehat P_1$ and $\widehat P_2$
connected by the blue line,
the purple line,
and the magenta--brown line.
Sheet~1 connects the magenta--brown line to the blue line,
then sheet~2 links the blue line to the purple line,
and finally sheet~3 goes from the purple line to the magenta--brown line.

\paragraph{Yellow line:}
The yellow line is on sheet~3.
It is the line where the bounding surface has minimal $\gamma_5 = 0$.
Setting $\gamma_5 = Q_5 = 0$ one obtains
\be
\varepsilon^2 = \frac{3 \left( 1 - \delta^2 \right)
  \left( 1 + \delta \right)}{32},
\ee
which defines the yellow line together with $\gamma_5 = 0$.
That line connects $\widehat P_1$ to $\widehat P_2$ just as the main lines.
It attains maximum
$\varepsilon = 1/3$
when $\delta = 1/3$.

\paragraph{Red line:}
This line connects $\widehat P_3$ to $\widehat P_4$
and then proceeds to $\widehat P_5$ through the equations
\bs
\ba
\delta &=& \frac{3 - 11 \varepsilon^2
  + 5\, \sqrt{\varepsilon^2 \left( 24 - 47 \varepsilon^2 \right)}}{27},
\\
\gamma_5 &=& \frac{54 - 504 \varepsilon^2 + 2\,359 \varepsilon^4
  - \sqrt{\varepsilon^2 \left( 24 - 47 \varepsilon^2 \right)^3}}{135}.
\ea
\es
Between $\widehat P_3$ and $\widehat P_4$ this line gives,
for each $\gamma_5$,
the point of maximum $\varepsilon$.

\paragraph{Orange line:}
This line is on sheet~3 and materializes the upper bound~\eqref{b2}.
It goes from $\widehat P_2$ to $\widehat P_3$ and is given by
\bs
\ba
\gamma_5 &=& \frac{9 \left( 1 - \delta \right)^2}{10},
\\
\varepsilon^2 &=& \frac{3 \left( 1 - \delta \right)}{8}.
\ea
\es

\begin{figure}[h!]
\begin{center}
\includegraphics[width=0.45\textwidth]{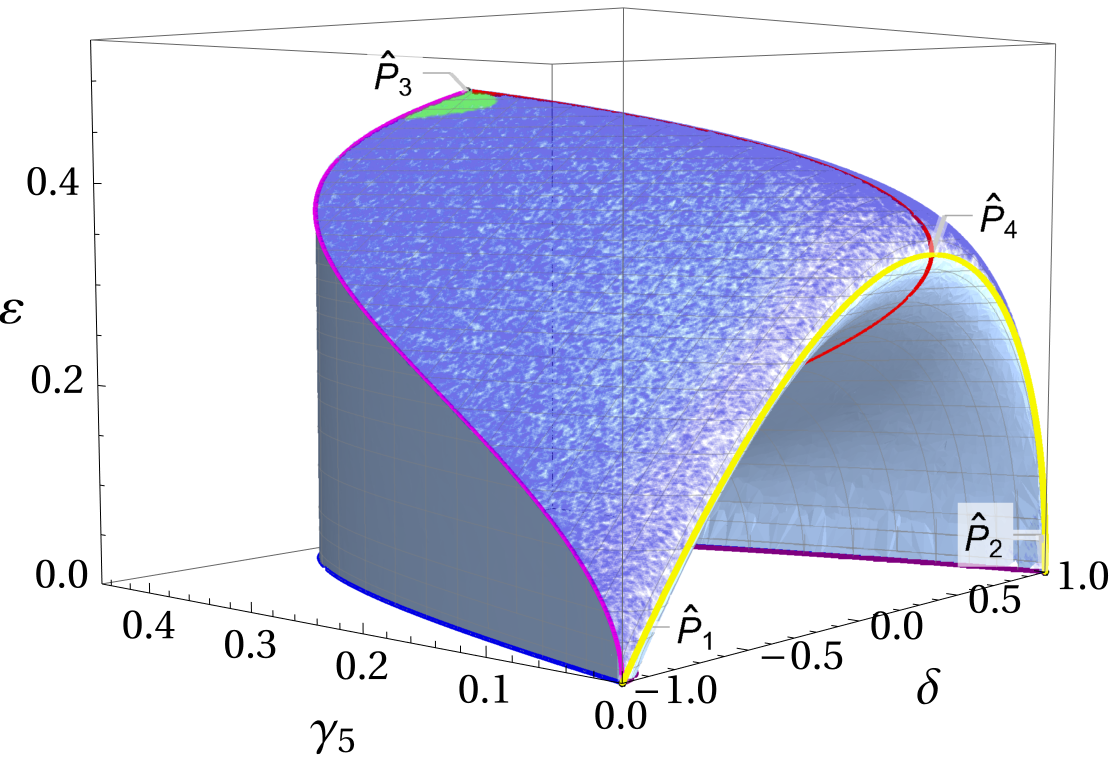}
\hspace{0.05\textwidth}
\includegraphics[width=0.46\textwidth]{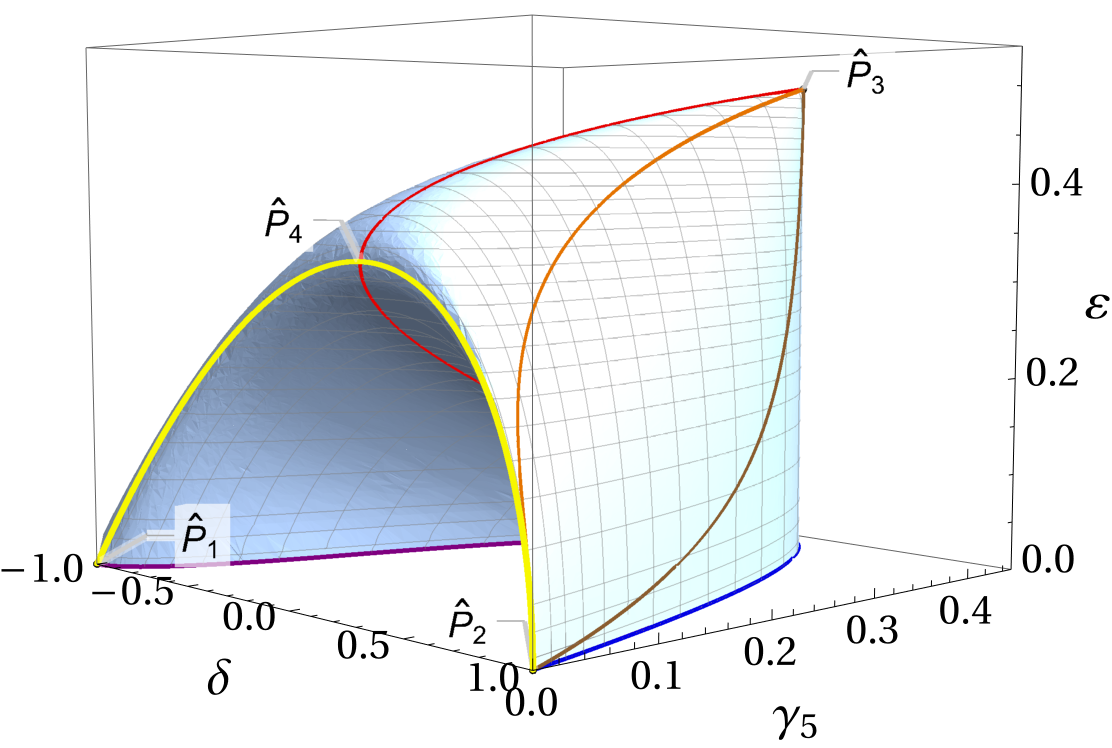}
\end{center}
  \caption{Two perspectives of the boundary of phase space.
    The points and lines displayed are defined in the text.
    The blue points in left panel indicate the convex part
    of the surface $Q_5=0$;
    green points correspond to the part of that surface
    that is neither concave nor convex.}
\label{fig:Y12K-3D}
\end{figure}

\begin{figure}[h!]
  \begin{center}
    \includegraphics[width=0.7\textwidth]{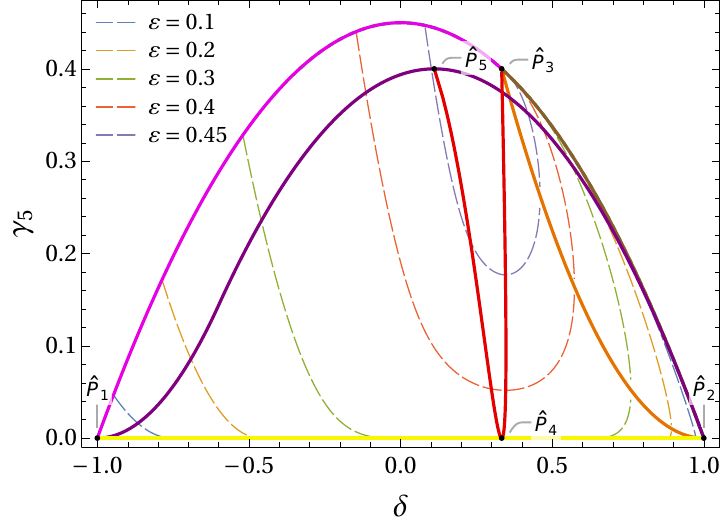}
  \end{center}
  \caption{The projection of phase space
    on the $\gamma_5$ \textit{vs.}\ $\delta$ plan.
    The points and lines displayed are defined in the text.
    The blue line coincides with the magenta--brown line in this projection.
    The iso-lines of constant
    $\varepsilon$
    on the upper part of sheet~3---the part of that sheet
    that connects the yellow line to the magenta--brown line---are marked
    by dashed lines in various colours.
       }
\label{fig:Y12K-proj1}
\end{figure}

\paragraph{Figures:}
Figure~\ref{fig:Y12K-3D} displays two perspectives of phase space.
Unfortunately,
on those perspectives the purple line is barely visible
and point $\widehat P_5$ is not visible at all,
therefore Fig.~\ref{fig:Y12K-proj1} shows the projection of phase space
on the $\gamma_5$ \textit{vs.}\ $\delta$ plan.

\paragraph{Convexity of sheet~3:}
The upward-facing part of sheet~3,
\textit{i.e.}\ the one with maximum $\varepsilon$,
is almost entirely convex,
except for a small region near the sheet's top
that is neither concave nor convex,
illustrated by the green area in
Fig.~\ref{fig:Y12K-3D}.
The red line lies close to,
but outside,
this green region.

\paragraph{Boundedness-from-below:}
The potential in Eq.~\eqref{c5}
is similar to the one in Eq.~\eqref{pot1},
so the BFB conditions~\eqref{hj22}--\eqref{hj24} once again hold,
just as the necessary BFB conditions~\eqref{29a},
\eqref{300},
and~\eqref{fj9d99};
one must also discard all the situations
where conditions~\eqref{imped} hold true.
Of course,
in the second Eq.~\eqref{pot26} one must make $\psi \to \nu$
and $\epsilon \to \varepsilon$.

\paragraph{Relevance of the lines:}
We have randomly generated $10^6$ sets of parameters
$\lambda_1, \ldots, \lambda_5, \nu$,
and we have explicitly minimized $V_4$ for each of these sets.
We have found that by applying conditions~\eqref{hj22}--\eqref{hj24}
just along the \emph{magenta, red, yellow, and orange lines}
allows one to discard the majority
(up to 99.95\%)
of the potentials that are not bounded from below.
In order to achieve exact agreement
with the results of the minimization of $V_4$,
the selected parameter sets
should be re-evaluated by applying conditions~\eqref{hj22}--\eqref{hj24}
for every possible $\left( \delta, \gamma_5, \varepsilon \right)$
on the upper part of sheet~3.
Since the convexity of phase space
is stronger than in the case $Y=1/2$ with reflection symmetry,
a larger number of points must be used for the
scan; we have used $10^5$ points 
$\left( \delta, \gamma_5, \varepsilon \right)$.\footnote{An
alternative approach would be to firstly minimize the quantity $\Delta$
in Eq.~\eqref{eq:Delta} over the upper part of sheet~3,
then to discard the points where $\Delta$ is negative,
and finally to apply condition~\eqref{uocppv2}.}

\paragraph{Procedure:} Our practical recommendation
for establishing the boundedness-from-below of  $V_4$
consists of the following steps.
\begin{enumerate}
\item The necessary conditions~\eqref{29a},
  \eqref{300},
  and~\eqref{fj9d99} are checked.
\item Cases satisfying conditions~\eqref{imped} are discarded.
\item Conditions~\eqref{hj22}–\eqref{hj24} are tested
  for $\aleph$ and $\beth$ defined through Eqs.~\eqref{aleph-beth}
  at the special points~\eqref{points_Y12K}.\footnote{It is worth noting that
  steps~1 and~2 by themselves alone allow the elimination of up to 91.45\%
  of the potentials violating the BFB conditions.
  Complemented with the scan over the special points,
  up to 97.76\% of the potentials violating the BFB conditions are discarded.}.
\item Conditions~\eqref{hj22}–\eqref{hj24} are tested
  for $\aleph$ and $\beth$ defined along the magenta,
  red,
  yellow,
  and orange lines.
\item For full precision,
  the remaining points must be
  scanned through $10^5$ points
  uniformly distributed over the whole upper part of surface
  $Q_5=0$.
\end{enumerate}

\newpage

\setcounter{equation}{0}
\renewcommand{\theequation}{D\arabic{equation}}

\section{Unitarity conditions}
\label{App:UNI}

For the model in Section~\ref{sec:3/2},
\textit{viz.}\ for the quartic potential
given by Eqs.~\eqref{V4} and~\eqref{VextrY32},
the unitarity conditions are
\bs
\ba
\left| \lambda_1 \right| &<& M,
\label{uni1a}
\\
\left| \lambda_2 \right| &<& M,
\label{uni1b}
\\
\left| \lambda_2 + 2 \lambda_5 \right| &<& M,
\label{uni1c}
\\
\left| \lambda_2 + \frac{3 \lambda_5}{5} \right| &<& M,
\label{uni1d}
\\
\left| \lambda_2 + \frac{9 \lambda_5}{5} \right| &<& M,
\label{uni1e}
\\
\left| \lambda_3 \right|
+ \frac{3}{4}\, \left| \lambda_4 \right| &<& M,
\label{uni2a}
\\
\left| \lambda_3 - \frac{5 \lambda_4}{4} \right| &<& M,
\label{uni2b}
\\
\left| 3 \lambda_1 + 5 \lambda_2 + 3 \lambda_5 \right|
+ \sqrt{\left( 3 \lambda_1 - 5 \lambda_2 - 3 \lambda_5 \right)^2
  + 32 \lambda_3^2} &<& 2 M,
\label{uni1h}
\\
\left| \lambda_1 + \lambda_2 - \frac{11}{5} \lambda_5 \right|
+ \sqrt{\left( \lambda_1 - \lambda_2 + \frac{11}{5} \lambda_5 \right)^2
  + 10 \lambda_4^2} &<& 2 M,
\label{uni2d}
\\
\left| \lambda_1 + \lambda_3 + \frac{5}{4} \lambda_4 \right|
+ \sqrt{\left( \lambda_1 - \lambda_3 - \frac{5}{4} \lambda_4 \right)^2
  + 24 \left| \psi \right|^2} &<& 2 M.
\label{uni2c}
\ea
\es

For the model in Section~\ref{sec:1/2},
\textit{viz.}\ for the quartic potential
given by Eqs.~\eqref{V4} and~\eqref{03kkfd},
the unitarity conditions are conditions~\eqref{uni1a},
\eqref{uni1b},
\eqref{uni1d},
\eqref{uni1e},
\eqref{uni1h},
\eqref{uni2d},
and
\bs
\ba
\left| \lambda_3 + \frac{3 \lambda_4}{4} \right| &<& M,
\\
\left| \lambda_3 - \frac{3 \lambda_4}{4} \right|
+ \frac{3}{\sqrt{10}}\, \left| \chi \right| &<& M,
\\
\left| \lambda_3 - \frac{5 \lambda_4}{4} \right| &<& M,
\\
\left| \lambda_3 + \frac{5 \lambda_4}{4} \right|
+ \sqrt{\frac{5}{2}}\, \left| \chi \right| &<& M,
\\
\left| \lambda_1 + \lambda_2 + 2 \lambda_5 \right|
+ \sqrt{\left( \lambda_1 - \lambda_2 - 2 \lambda_5 \right)^2
  + 4 \left| \chi \right|^2} &<& 2 M.
\ea
\es

For the model in Appendix~\ref{App:more},
\textit{viz.}\ for the quartic potential
given by Eqs.~\eqref{V4} and~\eqref{9v0f00f},
the unitarity conditions are conditons~\eqref{uni1a}--\eqref{uni2a},
\eqref{uni1h},
and
\bs
\ba
\left| \lambda_3 + \frac{5 \lambda_4}{4} \right| &<& M,
\\
\left| \lambda_1 + \lambda_3 - \frac{5}{4} \lambda_4 \right|
+ \sqrt{\left( \lambda_1 - \lambda_3 + \frac{5}{4} \lambda_4 \right)^2
  + 2 \left| \nu \right|^2} &<& 2 M.
\ea
\es
Furthermore,
the moduli of the three eigenvalues of the matrix
\be
\left( \begin{array}{ccc}
  \lambda_1 &
  -\sqrt{2} \left| \nu \right| &
  \sqrt{5/2}\, \lambda_4 \\
  -\sqrt{2} \left| \nu \right| &
  \lambda_3 + (5/4) \lambda_4 &
  0 \\
  \sqrt{5/2}\, \lambda_4 &
  0 &
  \lambda_2 - (11/5) \lambda_5
\end{array} \right)
\ee
must all be smaller than $M$.

When $V_{4,\mathrm{extra}} = 0$ the unitarity conditions
for all three models are identical,
\textit{viz.}\ conditions~\eqref{uni1a}--\eqref{uni1e},
\eqref{uni2a},
\eqref{uni1h},
\eqref{uni2d},
and
\ba
\left| \lambda_3 \right|
+ \frac{5}{4} \left| \lambda_4 \right| &<& M.
\ea
They coincide with the unitarity conditions
for the case $J = 3/2$ in Ref.~\cite{we}.

\end{appendix}

\end{document}